\documentclass[%
  reprint,
  notitlepage,
superscriptaddress,
preprintnumbers,
nofootinbib,
nobibnotes,
 amsmath,amssymb,
 aps,
 onecolumn,
]{revtex4-1}
\usepackage{graphicx}
\usepackage{natbib}
\usepackage{aas_macros,braket}
\usepackage{bm}
\usepackage{dcolumn}
\usepackage{multirow}
\begin{document}
\title{Testing statistics of the CMB B-mode polarization toward unambiguously establishing quantum fluctuation of vacuum}

\author{Maresuke Shiraishi}
%\email
\affiliation{%
Kavli Institute for the Physics and Mathematics of the Universe (Kavli IPMU, WPI), UTIAS, The University of Tokyo, Chiba, 277-8583, Japan
}

\author{Chiaki Hikage}
%\email
\affiliation{%
Kavli Institute for the Physics and Mathematics of the Universe (Kavli IPMU, WPI), UTIAS, The University of Tokyo, Chiba, 277-8583, Japan
}

\author{Ryo Namba}
%\email
\affiliation{%
Kavli Institute for the Physics and Mathematics of the Universe (Kavli IPMU, WPI), UTIAS, The University of Tokyo, Chiba, 277-8583, Japan
}

\author{Toshiya Namikawa}
%\email
\affiliation{%
Department of Physics, Stanford University, Stanford, CA 94305, USA
}
\affiliation{%
Kavli Institute for Particle Astrophysics and Cosmology, SLAC National Accelerator Laboratory, Menlo Park, CA 94025, USA  
}

\author{Masashi Hazumi}
%\email
\affiliation{%
Kavli Institute for the Physics and Mathematics of the Universe (Kavli IPMU, WPI), UTIAS, The University of Tokyo, Chiba, 277-8583, Japan
}
\affiliation{%
Institute of Particle and Nuclear Studies, KEK, 1-1 Oho, Tsukuba, Ibaraki 305-0801, Japan
}
\affiliation{%
SOKENDAI (The Graduate University for Advanced Studies), Hayama, Miura District, Kanagawa 240-0115, Japan
}
\affiliation{%
Institute of Space and Astronautical Studies (ISAS),
Japan Aerospace Exploration Agency (JAXA), Sagamihara, Kanagawa 252-5210, Japan}

\date{\today} \preprint{IPMU16-0086}

\begin{abstract}
The B-mode polarization in the cosmic microwave background (CMB) anisotropies at large angular scales provides a smoking-gun evidence for the primordial gravitational waves (GWs). It is often stated that a discovery of the GWs establishes the quantum fluctuation of vacuum during the cosmic inflation. Since the GWs could also be generated by source fields, however, we need to check if a sizable signal exists due to such source fields before reaching a firm conclusion when the B-mode is discovered. Source fields of particular types can generate non-Gaussianity (NG) in the GWs. Testing statistics of the B-mode is a powerful way of detecting such NG. As a concrete example, we show a model in which a gauge field sources chiral GWs via a pseudoscalar coupling, and forecast the detection significance at the future CMB satellite LiteBIRD. Effects of residual foregrounds and lensing B-mode are both taken into account. We find the B-mode bispectrum ``BBB'' is in particular sensitive to the source-field NG, which is detectable at LiteBIRD with a $> 3 \sigma$ significance. Therefore the search for the ``BBB'' will be indispensable toward unambiguously establishing quantum fluctuation of vacuum when the B-mode is discovered. We also introduced the Minkowski functional to detect the NGs. While we find that the Minkowski functional is less efficient than the harmonic-space bispectrum estimator, it still serves as a useful cross check. Finally, we also discuss the possibility of extracting clean information on parity violation of GWs, and new types of parity-violating observables induced by lensing.
\end{abstract}

%\pacs{98.80.Cq}

\maketitle

%%%%%%%%%%%%%%%%%%%%%%%%%%%%%%%%%%%%%%%%%%%%%%%%%%%%%%%%%%%%%%%%%%%%
\section{Introduction}
%%%%%%%%%%%%%%%%%%%%%%%%%%%%%%%%%%%%%%%%%%%%%%%%%%%%%%%%%%%%%%%%%%%%

Primordial inflation has become a compelling paradigm for the mechanism to provide the seeds of the current structures of our universe during the first tiny fraction of a second of the cosmic history. A number of recent observations have pinned down or placed stringent bounds on the cosmological parameters that describe the state of the present universe, with the inflationary predictions as the initial conditions. On the other hand, the physical state of the universe during inflation is not yet unveiled to a satisfactory degree; especially, the inflationary energy scale is an urgent target to hunt down. Moreover, inflation is likely to have occurred in the energy regime much higher than that of particle accelerators, and thus it can provide us a unique arena to study the physics in the energy scales otherwise beyond our reach.

Detection of the primordial B-mode of cosmic microwave background (CMB) fluctuations or equivalently the tensor mode of primordial perturbations, is often considered as a direct measure of the inflationary energy scale. This is due to the fact that the tensor modes, or gravitational waves (GWs), originated from vacuum fluctuations depend only on the background evolution of space, namely the Hubble parameter, and are independent of the dynamics of field contents that are present during inflation. Given the great importance, a number of projects are ongoing, such as the {\it Planck} satellite \cite{Ade:2015tva}, POLARBEAR \cite{Ade:2014afa}, BICEP2/Keck Array \cite{Array:2015xqh}, SPTpol \cite{Keisler:2015hfa} and ACTPol \cite{Naess:2014wtr} to see the first hint at the primordial GWs.
The current upper limit on the tensor-to-scalar ratio $r$, which is proportional to the primordial GW signal, is 0.07 at 95\% C.L. \cite{Array:2015xqh}.%
\footnote{See e.g. \cite{Bennett:2012zja,Barkats:2013jfa,Ade:2014xna,Ade:2015tva,Ade:2015xua} for the limits from the other B-mode data, and \cite{Ade:2013zuv,Planck:2013jfk,vanEngelen:2014zlh,Ade:2015xua,Ade:2015lrj} for the current limits from the temperature and/or E-mode polarization data. See e.g. \cite{Ade:2013hjl,Ade:2013gez,Ade:2014afa,Naess:2014wtr,Keisler:2015hfa,Ade:2015nch,Array:2016afx} for the measurements of lensed B-mode/lensing potential using polarization data.}
Proposals for measurements in 2020s such as LiteBIRD \cite{Hazumi:2012aa,Matsumura:2013aja,2016JLTP..tmp..169M}, PIXIE \cite{Kogut:2011xw} and CMB-S4 \cite{Abazajian:2013oma,Wu:2014hta} aim at the ambitious goal sensitivity of $\sigma(r) \sim 0.001$, where $\sigma(r)$ represents the
total uncertainty on $r$. 
 
After detection of any primordial B-modes, the next critical step is to reveal the origin of the primordial B-modes. Although multiple studies have assumed the primordial B-modes from quantum fluctuations during inflation, validity of this assumption should be tested by future observations. In fact, GWs are sourced by any energetically-relevant components during inflation, and particles abundantly produced in the inflationary era would inevitably imprint signatures in primordial B-modes. One obstacle to realize such models is that these particles could simultaneously generate significant density (scalar) perturbations inconsistent with current observational results. Therefore any consistent study of GWs from particle production must satisfy the constraints on the scalar perturbations.

The effect of particle production on GWs can be maximized as compared to that to scalar perturbations when (i) the production occurs in the sector whose dynamics is decoupled from the adiabatic direction (inflaton), and (ii) the produced particles have spin-1 or higher and are relativistic \cite{Cook:2011hg,Barnaby:2012xt}. Even if these requirements are fulfilled, it is still challenging to generate a detectable level of GWs, simply because the energy transferred to the scalar is significant and the CMB temperature observations place too stringent constraints \cite{Mirbabayi:2014jqa}. One of only few ways to circumvent this issue, extensively considered in \cite{Mirbabayi:2014jqa,Ferreira:2014zia}, is to have strongly scale-dependent spectra. As a concrete working example, we therefore focus on the model presented in \cite{Namba:2015gja}, which we hereafter call the pseudoscalar model, and analyze it in detail aiming for the ability of future observations.

In this model, particle production is localized in time, which in turn imprints a bump in the spectra of both tensor and scalar perturbations, enhancing only the modes that exit the horizon around the time of particle production. Due to this feature, an efficient energy transfer to the scalar mode takes place only for a short period and is suppressed compared to that to the tensor, evading the arguments in \cite{Mirbabayi:2014jqa,Ferreira:2014zia}. Therefore this is one of very few existing models that have an observationally viable parameter space for the generation of the B-mode of the origin uncorrelated with vacuum fluctuations.

GWs predicted in the pseudoscalar model have several unique features: a bump in the spectra, helical nature, and large non-Gaussianity (NG) in the B-modes. Local production during inflation leaves a peak in the power-/bispectra, and the production in this model is triggered by a parity-violating operator, resulting in helical GWs. It exhibits the distinctive features of scale dependence and parity, which are absent in the conventional mechanisms of GW production, with a significant signal-to-noise ratio (SNR). The parity violation leads to non-vanishing correlation in the B-mode auto bispectrum for the $\ell_1 + \ell_2 + \ell_3 = {\rm even}$ modes, since the parity-odd nature of B-mode leads only to excitations of $\ell_1 + \ell_2 + \ell_3 = {\rm odd}$ correlations in parity-conserving models \cite{Kamionkowski:2010rb,Shiraishi:2011st,Shiraishi:2014roa}. None of these features appears in the GWs originating from the ground-state vacuum, and testing them could play a significant role to reveal the origin of the primary B-modes. There are already massive studies on primordial NGs from temperature anisotropy data \cite{Komatsu:2008hk,Komatsu:2010fb,Bennett:2012zja,Ade:2013ydc,Ade:2015ava}. The Planck collaboration recently reported the first observational constraints from the E-mode polarization anisotropies \cite{Ade:2015ava}. On the other hand, primordial NGs have never been measured using B-mode maps so far.%
\footnote{See \cite{Shiraishi:2013kxa,Shiraishi:2013vha,Meerburg:2016ecv} for the sensitivity analyses on several types of primordial NGs in future B-mode surveys. See \cite{Shiraishi:2013wua,Shiraishi:2014ila,Ade:2015ava,Ade:2015cva} for the observational bounds on the GW bispectra from the temperature and/or E-mode polarization data.}
This paper motivates such a new attempt.

Our interest is to test the sensitivity of forthcoming data to the B-mode NG under a more {\it realistic} experimental setup. For this purpose, we take into account three types of uncertainties due to {\it (i)} instrumental features, {\it (ii)} astrophysical foregrounds and {\it (iii)} gravitational lensing. Concerning {\it (i)}, we assume a measurement in a next-generation satellite mission LiteBIRD \cite{Hazumi:2012aa,Matsumura:2013aja,2016JLTP..tmp..169M}. The main components of {\it (ii)} are dust emission and synchrotron radiation in our galaxy. These can be largely subtracted in current data analyses by using multiband data, so we assume that only $2\%$ level remains in CMB maps. The uncertainty {\it (iii)} means secondary B-modes converted from primary E-modes via gravitational lensing. The E-modes are mostly generated from primary scalar-mode perturbations, and the lensing B-modes become a contaminant in extracting primary GW or tensor-mode signal. The contribution from lensing could be partially subtracted with delensing techniques \cite{Kesden:2002ku,Knox:2002pe,Seljak:2003pn,Smith:2010gu,Namikawa:2015tjy}. In this work, we do not consider delensing analyses in the case with LiteBIRD data, because the lensing B-mode is small with respect to the primordial B-mode in the lowest multipole region and the sensitivity to the B-mode NG is not drastically improved by this procedure. On the other hand, in the noiseless case, to see the impact of the lensing contaminations on the estimate of the B-mode NG, we also compute the SNR assuming a perfect subtraction of lensing contributions. Including the effects of the above three contributions in the covariance matrix, this paper estimates {\it realistic} SNRs of B-mode NG.%
\footnote{We do not include effects of correlated (e.g. non-white) noise, NG beams and filtering as they depend highly on each experimental configuration. One should also consider correlations among different multipoles of observed B-modes that arise from the sky cut, which may result in leakage of the lensing bias in the $\ell_1 + \ell_2 + \ell_3 = \text{even}$ mode of the B-mode bispectrum into the $\ell_1 + \ell_2 + \ell_3 = \text{odd}$ mode and vise versa. The estimate of this leakage, however, requires a realistic numerical simulation, which is beyond the scope of this paper.}

We consider the following two ways to extract NGs from B-modes. The first one is the direct measurement of the CMB bispectrum in harmonic space. This gives an optimal result saturating the Cram\'er-Rao bound \cite{Komatsu:2003iq,Babich:2005en,Komatsu:2008hk,Liguori:2010hx,Komatsu:2010hc,Shiraishi:2014roa,Shiraishi:2014ila}. The second one is the measurement of the Minkowski functional (MF), characterizing the topology of CMB maps \cite{Mecke:1994ax,Schmalzing:1997aj,Matsubara:2003yt,Matsubara:2010te}. The MF includes integrated information of NGs, such as skewness and kurtosis, and this fact tends to reduce optimality compared with the direct bispectrum measurement. Nevertheless, the MF analysis is still valuable as an independent NG test, and it has been widely employed \cite{Hikage:2008gy,Hikage:2012bs,Ade:2013ydc,Ade:2015ava}. Our SNR estimations are done by computing the Fisher matrices of both the bispectrum and the MF. To extract clean primary tensor-mode signatures, we then limit our analysis of B-mode auto-bispectrum to the $\ell_1 + \ell_2 + \ell_3 = {\rm even}$ domain and shut out the scalar-mode lensing signal in the $\ell_1 + \ell_2 + \ell_3 = {\rm odd}$ domain. 
While in the bispectrum analysis one can do this by hand, in the MF analysis the $\ell_1 + \ell_2 + \ell_3 = {\rm odd}$ components are automatically prohibited due to the parity-conserving nature of skewness. In this light, the B-mode MF is a clean indicator of parity-violating NGs. We explicitly show this in this study.

The lensed B-mode NG signal originating from the scalar mode, which is treated as a bias in our analysis, is confined to the $\ell_1 + \ell_2 + \ell_3 = {\rm odd}$ domain due to parity invariance of the scalar mode. On the other hand, nonvanishing parity-violating correlations sourced by the primary tensor mode, e.g., the EB correlation and the cross power spectrum between B-mode polarization and the lensing potential, can produce nonvanishing lensed B-mode NG with $\ell_1 + \ell_2 + \ell_3 = {\rm even}$. We estimate this uninvestigated signal and verify that it contributes subdominantly to the SNR, compared with the primary B-mode bispectrum.

In addition to the B-mode NG, the TB and EB correlations are also distinctive CMB observables of the pseudoscalar model due to parity violation. However, sizable cosmic variances of the temperature and E-mode anisotropies reduce their sensitivities and the resultant SNRs fall below that of the BBB bispectrum. We therefore refrain from them and focus only on pure B-mode observables 
in this study.%
\footnote{See \cite{Saito:2007kt,Gerbino:2016mqb} for the limits on the chirality of primordial GWs from the TB and EB data.}

  This paper is organized as follows. In Sec.~\ref{sec:theory}, we summarize the pseudoscalar model \cite{Namba:2015gja}, which realizes sizable GW NG and visible B-mode bispectrum. In Secs.~\ref{sec:bis} and \ref{sec:MF}, we estimate SNRs of this B-mode bispectrum by means of the harmonic-space bispectrum analysis and the MF one, respectively. After discussing the distinguishability between the sourced GWs and the vacuum-induced ones in Sec.~\ref{sec:comparison}, we conclude this paper in Sec.~\ref{sec:conclusions}. In Appendix~\ref{appen:pow_BB}, we compute SNRs from the B-mode power spectrum. We analyze the B-mode bispectrum created through late-time gravitational lensing in Appendix~\ref{appen:bis_lens}. The power spectra involving the lensing potential, employed to calculate the lensed B-mode bispectrum in Appendix~\ref{appen:bis_lens}, are discussed in Appendix~\ref{appen:pow_lens}. The noise spectrum in LiteBIRD, used in our SNR estimations, is computed in Appendix~\ref{appen:noise}.

%%%%%%%%%%%%%%%%%%%%%%%%%%%%%%%%%%%%%%%%%%%%%%%%%%%%%%%%%%%%%%%%%%%%
\section{An inflationary model producing visible B-mode non-Gaussianity}\label{sec:theory}
%%%%%%%%%%%%%%%%%%%%%%%%%%%%%%%%%%%%%%%%%%%%%%%%%%%%%%%%%%%%%%%%%%%%

In this section, we introduce one of the few existing models that can produce a visible BBB auto-bispectrum at the scales relevant for the CMB observations, while respecting the bounds from the temperature anisotropies. The mechanism is based on particle production, and produced particles serve as a new source of GWs uncorrelated with the standard vacuum fluctuations. We here summarize the model briefly, and we refer interested readers to \cite{Namba:2015gja}, which first studied this specific model,%
\footnote{
The case where a scale-invariant spectrum of GWs in the same setup was first considered in \cite{Barnaby:2012xt}, and then the difficulty to evade the constraints from scalar perturbations was pointed out in \cite{Ferreira:2014zia}.
}
for technical details.

We assume that the background evolution of inflation is standard and is driven by a slowly-rolling inflaton, leading to a quasi de Sitter expansion. In the model of our interest, particle production is responsible for sourcing GWs. To avoid spin suppression, we consider production of vector (spin-$1$) fields. Since GWs are sourced by quadrupole moments, the mechanism is more effective with relativistic particles than massive (non-relativistic) ones \cite{Barnaby:2012xt}. While it is known \cite{Anber:2009ua} that a $U(1)$ gauge field is enhanced exponentially when it couples to an axion-like field, its production should necessarily be localized in time in order to minimize its effects to scalar perturbations \cite{Ferreira:2014zia,Namba:2015gja}. To realize such a production scenario, we assume an axion $\sigma$ to be a spectator field that has a negligible energy density compared to the inflaton. The coupling between an axion and a gauge field is fixed by symmetries (parity, shift, and gauge invariance) and takes the form
\begin{equation}
{\cal L}_{\rm int} = - \frac{\alpha}{4f} \, \sigma F_{\mu\nu} \tilde F^{\mu\nu} \; ,
\label{eq:Lint}
\end{equation}
where $\tilde F^{\mu\nu} \equiv \frac{\epsilon^{\mu\nu\rho\sigma}}{2 \sqrt{-g}} F_{\rho\sigma}$ is the dual of the field-strength tensor $F_{\mu\nu}$ of a $U(1)$ gauge field $A_\mu$, $f$ is the axion decay constant, and $\alpha$ is the coupling constant. This interaction modifies the dispersion relation of $A_\mu$ and leads to copious production provided that $\dot\sigma \ne 0$, where dot denotes derivative with respect to the cosmic time $t$. Note that a non-zero vacuum expectation value of $\dot\sigma$ spontaneously breaks parity. The equation of motion for $A_\mu$ in the flat FLRW background reads
\begin{equation}
\left( \frac{\partial^2}{\partial\tau^2} + k^2 \mp 2 a k H \xi \right) A_\pm = 0 \; ,
\label{eq:eom-Apm}
\end{equation}
where $A_\pm$ are circularly polarized states of the gauge field in the Fourier space, $\tau$ is the conformal time, $a$ is the scale factor, $H \equiv \dot{a}/a$ is the Hubble parameter, and $\xi \equiv \alpha \dot\sigma / (2 f H)$ is the effective coupling strength. The $\mp$ sign reflects the parity-violating nature of the operator \eqref{eq:Lint} (when $\dot\sigma \ne 0$). As is clear from the equation, the last term in the parentheses of Eq.~\eqref{eq:eom-Apm} dominates over the $k^2$ term when $k/(aH) < 2 \vert\xi\vert$, and one of the polarization states experiences tachyonic instability (the other state is not produced). This therefore leads to an exponential production of a helical gauge field. In order to realize localized production, we take a simple and natural potential for an axion field, given as
\begin{equation}
V(\sigma) = \Lambda^4 \left( 1 + \cos \frac{\sigma}{f} \right) \; ,
\label{eq:V-sig}
\end{equation}
where $\Lambda$ is a mass parameter. This potential allows a very small $\dot\sigma$ at both early and late times and a relatively large $\dot\sigma$ around $\sigma / f = \pi / 2$. Since the production of the gauge field is exponential in $\dot\sigma$, even a small increase within a slow-roll regime of $\sigma$ can drastically enhance the efficiency of production. Assuming an almost de Sitter expansion and slow roll of $\sigma$, the effective coupling strength is given as \cite{Namba:2015gja},
\begin{equation}
\xi = \frac{\xi_*}{\cosh \left[ H \delta \left( t - t_* \right) \right]} \; ,
\label{eq:xi-sol}
\end{equation}
where the subscript $*$ denotes the value at the time when $\sigma / f = \pi / 2$, and $\delta \equiv \Lambda^4 / (3 H^2 f^2)$. Thus $\xi$ is peaked at time $t=t_*$, and its peak value is $\xi_* \equiv \alpha \delta / 2$. The width of this bump feature is controlled by $\delta^{-1}$, and thus a larger $\delta$ leads to a sharper and higher bump. The slow-roll condition for $\sigma$, i.e. $\ddot\sigma \ll 3 H \dot\sigma$, rewrites as $\delta \ll 3$. 
Without loss of generality, we hereafter take $\xi_* > 0$. In this case, Eq.~\eqref{eq:eom-Apm} tells us that only the $+$ mode of the gauge field receives enhancement, and using Eq.~\eqref{eq:xi-sol}, one can find a semi-analytical solution for $A_+$ \cite{Namba:2015gja}, which we use for our analysis.

The produced gauge quanta in turn source both scalar and tensor perturbations. 
In the spatially flat gauge, since the total energy density is dominated by that of inflaton, the curvature (scalar) perturbation can be approximated as
\begin{equation}
\zeta \cong - \frac{H}{\dot\phi} \, \delta\varphi \; ,
\label{eq:def-zeta}
\end{equation}
where $\delta\varphi$ is the perturbation of inflaton around its vacuum expectation value $\phi$. The dominant contribution to $\zeta$ from the produced gauge field is in fact through the mass mixing term of $\delta\varphi$ with the spectator perturbation $\delta\sigma$, which arises gravitationally by integrating out non-dynamical variables. Collecting dominant contributions and neglecting slow-roll suppressed terms, the equation of motion for $\delta\varphi$ reads \cite{Namba:2015gja},
\begin{equation}
\left( \frac{\partial^2}{\partial\tau^2} + k^2 - \frac{a''}{a} \right) \left( a \, \delta\varphi \right) \simeq a^2 \, \frac{3 \, \dot\phi \dot\sigma}{M_p^2} \left( a \, \delta\sigma \right) \; ,
\label{eq:eom-dphi}
\end{equation}
in the Fourier space. The effects of the gauge field are encoded through $\delta\sigma$, whose equation of motion is
\begin{equation}
\left( \frac{\partial^2}{\partial\tau^2} + k^2 - \frac{a''}{a} \right) \left( a \, \delta\sigma \right) \simeq a^3 \frac{\alpha}{f} \int \frac{d^3x}{(2\pi)^{3/2}} \, {\rm e}^{-i {\bf k} \cdot {\bf x}} \, {\bf E} ( \tau , {\bf x}) \cdot {\bf B} ( \tau , {\bf x})
\; ,
\label{eq:eom-dsig}
\end{equation}
where ${\bf E}$ and ${\bf B}$ are ``electric'' and ``magnetic'' fields associated with $A_\mu$. The part of the solution to Eq.~\eqref{eq:eom-dphi} that is sourced by $A_\mu$ via $\delta\sigma$ can be obtained by the method of Green function \cite{Namba:2015gja}, and it converts to sourced curvature perturbation, denoted by $\zeta^{(1)}$, through the relation \eqref{eq:def-zeta}. This sourcing process is schematically depicted as $A_\mu + A_\mu \rightarrow \delta\sigma \rightarrow \delta\varphi \sim \zeta$.
For tensor perturbations, on the other hand, we decompose the traceless and transverse part of the spatial metric, $\delta g_{ij}^{TT} / a^2 \equiv h_{ij}$, into the circularly polarized states $\lambda = \pm$, as $h_\lambda(\tau, {\bf k}) = \Pi_\lambda^{ij}(\hat{k}) \, h_{ij}(\tau, {\bf k}) $, where $\Pi^{ij}_\lambda$ is the circular polarization tensor, obeying $\hat{k}_i \Pi^{ij}_\lambda(\hat{k}) = \Pi^{ii}_\lambda(\hat{k}) = 0$, $\Pi^{ij *}_\lambda(\hat{k}) \, \Pi_{ij}^{\lambda'}(\hat{k}) = \delta_{\lambda \lambda'}$ and $\Pi^{ij *}_\lambda(\hat{k}) = \Pi^{ij}_{-\lambda}(\hat{k}) = \Pi^{ij}_\lambda(-\hat{k})$. Tensor perturbations are sourced by the traceless and transverse part of the energy-momentum tensor, and the equation of motion for $h_\pm$ reads \cite{Namba:2015gja},
\begin{equation}
  \left( \frac{\partial^2}{\partial\tau^2} + k^2 - \frac{a''}{a} \right) \left( a \, h_\pm \right) = -\frac{2 a^3}{M_p^2} \, \Pi^{ij}_\pm ( \hat{k} ) \,
  \int \frac{d^3x}{(2\pi)^{3/2}} \, {\rm e}^{-i {\bf k} \cdot {\bf x}} 
\left[  E_i  E_j + B_i  B_j  \right]
\; .
\label{eq:eom-hpm}
\end{equation}
We can then solve Eq.~\eqref{eq:eom-hpm} for the sourced part of $h_\lambda$, using again the Green function.

The total curvature and tensor perturbations are the sum of vacuum and sourced modes,
\begin{equation}
  \zeta = \zeta^{(0)} + \zeta^{(1)} \; , \quad 
  h_{\pm} = h_{\pm}^{(0)} + h_{\pm}^{(1)} \; ,
\label{eq:zeta-h-sum}
\end{equation}
where the superscripts $(0)$ and $(1)$ denote vacuum and sourced modes, respectively. We define their (primary) power spectra in the standard manner,
\begin{eqnarray}
  {\cal P}_\zeta ( k_1 ) \, \delta^{(3)} \left( {\bf k}_1 + {\bf k}_2 \right)
  &\equiv& \frac{k_1^3}{2\pi^2} \Braket{\zeta ( {\bf k}_1) \zeta ({\bf k}_2)} \; ,
  \nonumber\\
    {\cal P}_{\lambda_1} ( k_1 ) \, \delta_{\lambda_1 \lambda_2} \, \delta^{(3)} \left( {\bf k}_1 + {\bf k}_2 \right)
    &\equiv& \frac{k_1^3}{2\pi^2} \Braket{h_{\lambda_1} ({\bf k}_1 ) h_{\lambda_2} ({\bf k}_2)} \; .
\label{eq:def-power}
\end{eqnarray}
Since $\zeta^{(0)}$ and $h_\pm^{(0)}$ are composed of creation/annihilation operators of $\delta\varphi$ and $h_\pm$, respectively, while the sourced modes come from those of $A_\pm$, no cross correlations between the vacuum and sourced modes have connected parts. Thus these uncorrelated contributions simply add up in the total power spectra, i.e.
\begin{equation}
{\cal P}_\zeta = {\cal P}_\zeta^{(0)} + {\cal P}_\zeta^{(1)} \; , \quad
{\cal P}_h = \sum_{\lambda=\pm} \left[ {\cal P}_\lambda^{(0)} + {\cal P}_\lambda^{(1)} \right] \simeq {\cal P}_h^{(0)} + {\cal P}_+^{(1)} \; ,
\label{eq:power-sum}
\end{equation}
where ${\cal P}_\zeta^{(0)} = H^2 / (8 \pi^2 \epsilon_\phi M_p^2 )$ and ${\cal P}_h^{(0)} \equiv {\cal P}_+^{(0)} + {\cal P}_-^{(0)} = 16 \epsilon_\phi {\cal P}_\zeta^{(0)}$ with $\epsilon_\phi$ denoting the slow-roll parameter for inflaton. Due to parity-violating nature of Eq.~\eqref{eq:Lint}, in the case of $\xi>0$, only the $A_+$ modes are enhanced, and consequently they source only the $\lambda = +$ helicity state of the tensor perturbations. Therefore it suffices to consider ${\cal P}_+^{(1)}$ in the sourced-mode contributions (if $\xi<0$, only $\lambda = -$ modes are produced). In contrast, the vacuum modes do not distinguish between ${\cal P}_+^{(0)}$ and ${\cal P}_-^{(0)}$.
Similarly, bispectra consist of uncorrelated vacuum and sourced contributions. As it is well known that the vacuum bispectra are unobservably small \cite{Acquaviva:2002ud,Maldacena:2002vr}, only the sourced modes are of our interest. We thus define the (primary) bispectra as
\begin{eqnarray}
  {\cal B}^\zeta_{k_1  k_2  k_3} \, \delta^{(3)} \left( {\bf k}_1 +  {\bf k}_2 + {\bf k}_3 \right)
  &\equiv& \Braket{\zeta^{(1)} ( {\bf k}_1 ) \zeta^{(1)}( {\bf k}_2 ) \zeta^{(1)} ( {\bf k}_3 )} \; ,
\nonumber\\
{\cal B}^{\lambda_1 \lambda_2 \lambda_3}_{k_1 k_2 k_3} \, \delta^{(3)} \left( {\bf k}_1 +  {\bf k}_2 + {\bf k}_3 \right)
&\equiv& \Braket{h_{\lambda_1}^{(1)} ( {\bf k}_1 )  h_{\lambda_2}^{(1)} ( {\bf k}_2 )  h_{\lambda_3}^{(1)} ( {\bf k}_3 ) } \; .
\label{eq:def-bispec}
\end{eqnarray}
In the same manner as the power spectrum, the parity-violating nature yields ${\cal B}^{+++} \gg {\cal B}^{++-} \, , {\cal B}^{+--} \, , {\cal B}^{---}$.

All non-standard features of the scalar and tensor perturbations in this model are encoded in the modes sourced by the gauge field, namely ${\cal P}_\zeta^{(1)}$, ${\cal P}_+^{(1)}$, ${\cal B}^\zeta$ and ${\cal B}^{+++}$. The production of gauge quanta is controlled by the effective coupling $\xi$, and as discussed below Eq.~\eqref{eq:xi-sol}, $\xi$ has a bump in the time direction. Since enhancement of modes is affected by the value of $\xi$ at the time of horizon crossing, the gauge-field production is localized in momentum space around the modes that cross horizon when $\xi$ has a peak value. Hence, there are $3$ parameters that determine spectral features of scalar and tensor perturbations: $\delta$ fixes the width of the bump in the spectra, $\xi_*$ controls its height, and $k_* \equiv a H \vert_{t=t_*}$ determines its location, corresponding to the mode that crosses horizon at the time of $\xi = \xi_*$. Since correlations occur among the modes with $k \sim k_*$, the power spectra can be enhanced at $k \sim k_*$ and the bispectra \eqref{eq:def-bispec} are peaked at equilateral configurations with $k_1 \sim k_2 \sim k_3 \sim k_*$, boosting the CMB correlators at the corresponding multipoles $\ell \sim k_* \tau_0$, with $\tau_0$ denoting the present horizon scale (see Figs.~\ref{fig:blllBBB} and \ref{fig:ClBB_pseudo}). In the following analysis, to see the impacts of these peaks on several CMB scales, we work on three specific values of $k_*$ as $7 \times 10^{-5} \, {\rm Mpc}^{-1}$, $5\times 10^{-4} \, {\rm Mpc}^{-1}$ and $5 \times 10^{-3} \, {\rm Mpc}^{-1}$.

The power spectra \eqref{eq:power-sum} are the superposition of (uncorrelated) vacuum and sourced modes, and thus the bump feature appears on top of the standard quasi scale-invariant spectra. If this feature is dominant, the spectra are highly scale-dependent. Such scale dependence should not be overwhelming in the scalar power spectrum, to be consistent with the observed TT correlation. Our goal is to search for a parameter region where the B-modes/GWs due to particle production are visible while the temperature fluctuations are within the observed bounds. As can be seen in Eq.~\eqref{eq:eom-dphi}, the scalar perturbation is sourced through $\delta\sigma$, and thus the sourcing is effective only for the duration when $\dot\sigma$ is sufficiently large. Therefore for a fixed value of $\xi_*$, the constraints from temperature fluctuations are weaker for a larger $\delta$ (a sharper bump). To maximize the sourcing of tensor perturbations, we thus take and fix a rather large value, $\delta = 0.5$, for our entire analysis in the following sections. Concerning $\xi_*$, we choose the maximum values allowed within $\sim 2 \sigma$ deviation from the WMAP best-fit for $\delta = 0.5$, reading $5.3$, $5.1$ and $4.9$ for $k_*  = 7 \times 10^{-5} \ \text{Mpc}^{-1}$, $5 \times 10^{-4} \ \text{Mpc}^{-1}$, and $5 \times 10^{-3} \ \text{Mpc}^{-1}$, respectively \cite{Namba:2015gja}.%
\footnote{
In the recent paper \cite{Peloso:2016gqs}, it has been verified that the calculations with these choices of parameters are under perturbative control.
}
In this case, although the TTT bispectrum is invisibly small, the BBB bispectrum is visibly amplified \cite{Namba:2015gja}. In the following sections, we shall investigate the detectability of this BBB bispectrum under lensing bias and realistic experimental settings. See also Appendix~\ref{appen:pow_BB} for the detectability analysis of the BB power spectrum.

Another unique feature of sourced spectra in this model is parity violation, originating from the axion-gauge interaction \eqref{eq:Lint}. This results in power asymmetry between the two helicity states of tensor perturbations, as seen in Eq.~\eqref{eq:power-sum}. This helical nature of tensor is observationally relevant in the form of non-zero TB and EB correlations \cite{Lue:1998mq} and non-vanishing BBB auto-bispectra for even $\ell_1 + \ell_2 + \ell_3$ \cite{Kamionkowski:2010rb,Shiraishi:2011st}. This parity-violating nature, together with the bump feature, is the smoking gun for the signals from this model.

%%%%%%%%%%%%%%%%%%%%%%%%%%%%%%%%%%%%%%%%%%%%%%%%%%%%%%%%%%%%%%%%%%%%%
\section{B-mode bispectrum}\label{sec:bis}

\begin{figure}[t]
  \begin{tabular}{cc}
    \begin{minipage}{0.5\hsize}
  \begin{center}
    \includegraphics[width=1\textwidth]{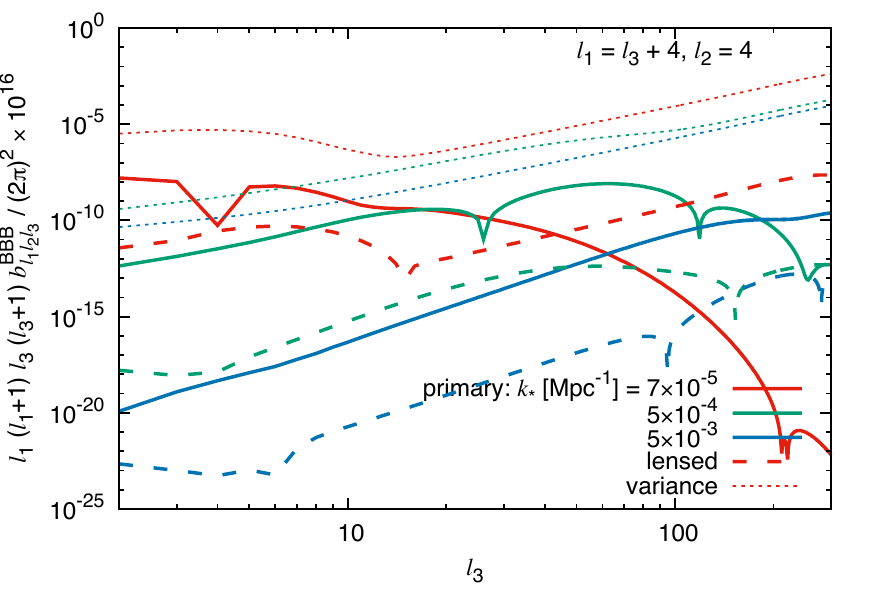}
  \end{center}
\end{minipage}
\begin{minipage}{0.5\hsize}
  \begin{center}
    \includegraphics[width=1\textwidth]{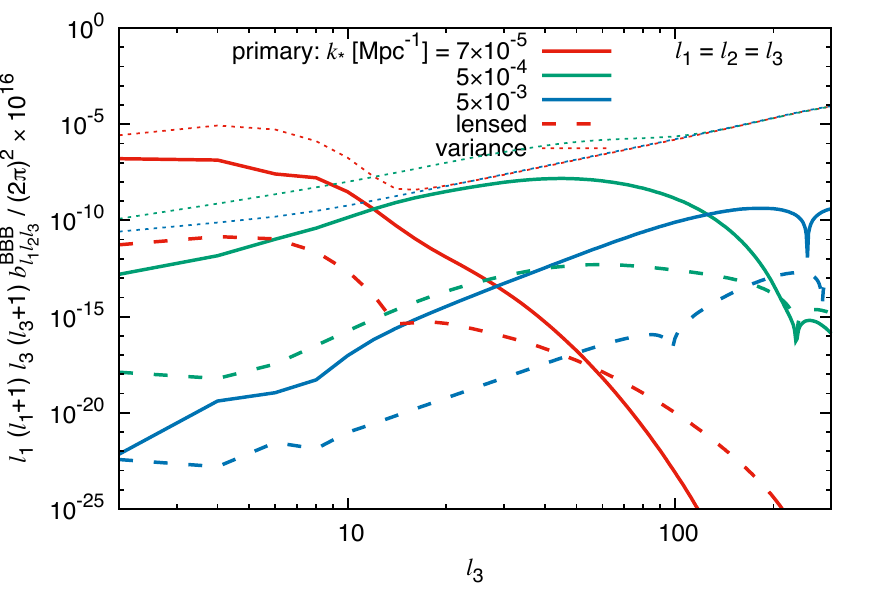}
  \end{center}
\end{minipage}
\end{tabular}
    \caption{Squeezed-limit (left panel) and equilateral-limit (right panel) signals of the primary B-mode bispectra \eqref{eq:BBB_prim} (solid lines) and the lensed B-mode bispectra due to the tensor mode \eqref{eq:BBB_lens} (dashed lines) in the $\ell_1 + \ell_2 + \ell_3 = {\rm even}$ sector. For comparison, we also show the level of the cosmic variance $\sigma(B_{\ell_1 \ell_2 \ell_3}^{BBB}) \simeq \sqrt{6 \tilde{C}_{\ell_1}^{BB} \tilde{C}_{\ell_2}^{BB} \tilde{C}_{\ell_3}^{BB}}$ (dotted lines).}
\label{fig:blllBBB}
\end{figure}

As analyzed in Appendix~\ref{appen:pow_BB} and shown in Fig.~\ref{fig:SNBB}, the BB power spectra for $k_* = 7 \times 10^{-5} \, {\rm Mpc^{-1}}$ and $5 \times 10^{-4} \, {\rm Mpc^{-1}}$ can be measured with very high significance in a LiteBIRD-like measurement. We now examine the possibility to detect the B-mode NG by an optimal CMB bispectrum estimator. 

Moving to harmonic space via $B(\hat{n}) = \sum_{\ell m} a_{\ell m}^B Y_{\ell m}(\hat{n})$, the CMB B-mode fluctuations (equivalent to Eq.~\eqref{eq:almB_ori}) are expressed as \cite{Zaldarriaga:1996xe,Hu:1997hp,Shiraishi:2010sm,Shiraishi:2010kd}
\begin{eqnarray}
  a_{\ell m}^{B} &=& 4\pi i^\ell  \int \frac{d^3 k}{(2\pi)^{3/2}}
 {\cal T}_{B,\ell}^{(t)}(k) \left[ h_+({\bf k})  -  (-1)^{\ell} h_-(-{\bf k})   \right] {}_{-2}Y_{\ell m}^*(\hat{k}) ~,  
\end{eqnarray}
where ${\cal T}_{B,\ell}^{(t)}(k)$ is the B-mode radiation transfer function due to the tensor-mode perturbation. This and Eq.~\eqref{eq:def-bispec} result in the BBB bispectrum 
\begin{eqnarray}
  \Braket{a_{\ell_1 m_1}^B a_{\ell_2 m_2}^B a_{\ell_3 m_3}^B}
  &=& \left[\prod_{n=1}^3 4\pi i^{\ell_n}  \int \frac{d^3 k_n}{(2\pi)^{3/2}}
    {\cal T}_{B,\ell_n}^{(t)}(k_n) {}_{-2}Y_{\ell_n m_n}^*(\hat{k}_n) \right]
\delta^{(3)}({\bf k}_1 + {\bf k}_2 + {\bf k}_3)
  \nonumber \\ 
  &&\times \left[ {\cal B}^{+++}_{k_1 k_2 k_3}
    - (-1)^{\ell_1 + \ell_2 + \ell_3}{\cal B}^{---}_{k_1 k_2 k_3} \right. \nonumber \\
  &&\left. \quad  
  + (-1)^{\ell_2+\ell_3}\left( {\cal B}^{+--}_{k_1 k_2 k_3}
  - (-1)^{\ell_1+\ell_2+\ell_3}{\cal B}^{-++}_{k_1 k_2 k_3} \right) \right. \nonumber \\
  &&\left.\quad 
  + (-1)^{\ell_1+\ell_3}\left( {\cal B}^{-+-}_{k_1 k_2 k_3}
  - (-1)^{\ell_1+\ell_2+\ell_3}{\cal B}^{+-+}_{k_1 k_2 k_3} \right) \right. \nonumber \\
  &&\left.\quad 
  + (-1)^{\ell_1+\ell_2}\left({\cal B}^{--+}_{k_1 k_2 k_3}
  - (-1)^{\ell_1+\ell_2+\ell_3}{\cal B}^{++-}_{k_1 k_2 k_3} \right) \right] ~. \label{eq:BBB_prim}
\end{eqnarray} 
It is obvious from this expression that if parity is preserved; namely, ${\cal B}^{\lambda_1 \lambda_2 \lambda_3} = {\cal B}^{-\lambda_1 -\lambda_2 -\lambda_3}$ holds, the $\ell_1 + \ell_2 + \ell_3 = {\rm even}$ components completely vanish. However, the pseudoscalar model realizes chiral GW bispectrum and ${\cal B}^{+++}$ is maximally enhanced, and thus those components give non-zero contributions. Consequently,  non-vanishing signals appear not only in $\ell_1 + \ell_2 + \ell_3 = {\rm odd}$ but also in $\ell_1 + \ell_2 + \ell_3 = {\rm even}$, and one can extract clean signatures of parity violation by looking at the latter \cite{Kamionkowski:2010rb,Shiraishi:2011st,Shiraishi:2012sn,Shiraishi:2013kxa}. Since rotational symmetry is respected, we can decompose the BBB bispectrum as 
\begin{eqnarray}
  \Braket{a_{\ell_1 m_1}^B a_{\ell_2 m_2}^B a_{\ell_3 m_3}^B}
  = B_{\ell_1 \ell_2 \ell_3}^{BBB}
   \left(
  \begin{array}{ccc}
  \ell_1 & \ell_2 & \ell_3 \\
  m_1 & m_2 & m_3 
  \end{array}
\right) ~, \label{eq:BBB_def}
\end{eqnarray}
where $B_{\ell_1 \ell_2 \ell_3}^{BBB}$ is called the angle-averaged bispectrum. Figure~\ref{fig:blllBBB} describes the squeezed and equilateral components of the parity-even reduced bispectra $b_{\ell_1 \ell_2 \ell_3}^{BBB} \equiv B_{\ell_1 \ell_2 \ell_3}^{BBB} / h_{\ell_1 \ell_2 \ell_3}$, where $h_{\ell_1 \ell_2 \ell_3} \equiv h_{\ell_1 \ell_2 \ell_3}^{0~0~0}$ and
\begin{eqnarray}
  h_{l_1 l_2 l_3}^{s_1 s_2 s_3} 
&\equiv& \sqrt{\frac{(2 l_1 + 1)(2 l_2 + 1)(2 l_3 + 1)}{4 \pi}}
\left(
  \begin{array}{ccc}
  l_1 & l_2 & l_3 \\
  s_1 & s_2 & s_3
  \end{array}
 \right) ~, \label{eq:hsym}
\end{eqnarray}
showing the existence of the expected peaks at $\ell \sim k_* \tau_0$.

\begin{figure}[t!]
  \begin{tabular}{c}
    \begin{minipage}{1.0\hsize}
  \begin{center}
    \includegraphics[width = 0.85\textwidth]{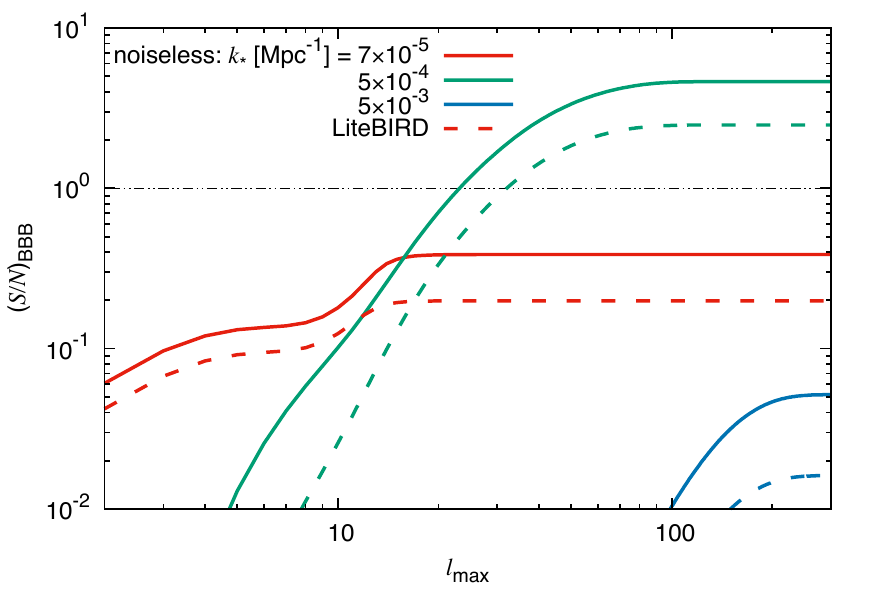}
  \end{center}
\end{minipage}
\end{tabular}
  \caption{SNRs of the primary B-mode bispectra for three different $k_*$ values in the pseudoscalar model, assuming a noiseless all-sky survey: $C_{\ell, \rm dat}^{BB} = C_{\ell, \rm prim}^{BB} + C_{\ell, \rm lens}^{BB}$ and $f_{\rm sky} = 1$ (solid lines), and LiteBIRD: $C_{\ell, \rm dat}^{BB} = C_{\ell, \rm prim}^{BB} + C_{\ell, \rm lens}^{BB} + N_{\ell, \rm LiteBIRD}^{BB}$ and $f_{\rm sky} = 0.5$ (dashed lines). We here limit the analysis to $\ell_1 + \ell_2 + \ell_3 = {\rm even}$. 
  }\label{fig:SNBBB_even} 
\end{figure}

For simplicity, we shall ignore off-diagonal components and any NG effects in the inverse of the bispectrum covariance.%
\footnote{Noise inhomogeneities, beam asymmetry, etc could generate off diagonal covariance, and the total signal to noise presented in this work would be degraded. The evaluation of this degradation however requires a realistic simulations based on LiteBIRD scan strategy, beam map simulation, etc, and will be explored in our future work.}
The inverse of the covariance is then given by an isotropic angular power spectrum $C_{\ell, \rm dat}^{BB} \equiv C_{\ell, \rm prim}^{BB} + C_{\ell, \rm lens}^{BB} + N_\ell^{BB} \equiv \tilde{C}_{\ell}^{BB} + N_\ell^{BB}$, with $C_{\ell, \rm prim}^{BB}$, $C_{\ell, \rm lens}^{BB}$ and $N_\ell^{BB}$ denoting the primary tensor-mode signal, the additional lensed signal converted from unlensed E-mode polarization and the lensing potential $\phi$, and the noise spectrum due to instrumental uncertainties plus residual foreground, respectively (see Fig.~\ref{fig:ClBB_pseudo} for their magnitude relations). This simplifies the expression of SNR to  
\begin{eqnarray}
  \left( \frac{S}{N} \right)_{BBB}^2 = \sum_{\ell_1, \ell_2, \ell_3 = 2}^{\ell_{\rm max}} \frac{|B_{\ell_1 \ell_2 \ell_3}^{BBB}|^2}{6 C_{\ell_1, \rm dat}^{BB} C_{\ell_2, \rm dat}^{BB} C_{\ell_3, \rm dat}^{BB}} ~. \label{eq:SNBBB}
\end{eqnarray}
In this expression, both the lensed spectrum and the noise spectrum contaminate the primary signal. The contributions of lensing are highly NG if going to higher $\ell$. At $\ell \lesssim 300$, the NG of the lensing B-modes is not significant (e.g. covariance of the B-mode power spectrum is at percent level \cite{BenoitLevy:2012va,Namikawa:2015tba}), and we set $\ell_{\rm max} = 300$ in our calculation. We assume a realistic NG measurement in LiteBIRD and therefore consider the bias due to residual galactic foreground with $2\%$-level magnitude in CMB maps, in addition to the bias coming from instrumental noise. All these biases are expressed as the power spectrum $N_\ell^{BB}$, and we disregard higher-order contributions for simplicity. The details of $N_\ell^{BB}$ are summarized in Appendix~\ref{appen:noise}. 

The bispectrum in the numerator of Eq.~\eqref{eq:SNBBB} should involve the late-time lensed signal converted from the primary scalar and tensor perturbations, in addition to the pure primary signal. Because of the spin-0 nature of the scalar mode, the scalar-mode lensed BBB is confined to the $\ell_1 + \ell_2 + \ell_3 = {\rm odd}$ domain. Our aim is to extract primordial GW information, thus, we limit our SNR computation to the parity-even configuration, $\ell_1 + \ell_2 + \ell_3 = {\rm even}$, and ignore the other. On the other hand, non-vanishing parity-even signal can also arise from the tensor-mode lensing, which is evaluated as Eq.~\eqref{eq:BBB_lens}. Adding this, in principle, improves the detectability of the tensor-mode signal. However, as seen in Fig.~\ref{fig:blllBBB}, this is always much smaller than the cosmic variance $\sigma(B_{\ell_1 \ell_2 \ell_3}^{BBB}) \simeq \sqrt{6 \tilde{C}_{\ell_1}^{BB} \tilde{C}_{\ell_2}^{BB} \tilde{C}_{\ell_3}^{BB}}$ and hence a negligible component in the SNR, despite exceeding the primary BBB for large $\ell$. One can see details of the lensed bispectrum in Appendix~\ref{appen:bis_lens}.

Figure~\ref{fig:SNBBB_even} describes SNRs for the three different $k_*$, as a function of $\ell_{\rm max}$, showing that only the $k_* = 5 \times 10^{-4} \, {\rm Mpc^{-1}}$ case is visible (i.e., the SNR can exceed unity) both in a noiseless full-sky survey ($C_{\ell, \rm dat}^{BB} = C_{\ell, \rm prim}^{BB} + C_{\ell, \rm lens}^{BB}$ and $f_{\rm sky} = 1$) and in LiteBIRD ($C_{\ell, \rm dat}^{BB} = C_{\ell, \rm prim}^{BB} + C_{\ell, \rm lens}^{BB} + N_{\ell, \rm LiteBIRD}^{BB}$ and $f_{\rm sky} = 0.5$).%
\footnote{Ref.~\cite{Namba:2015gja} showed that all the three $k_*$ cases are detectable in a noiseless full-sky survey if removing the lensing bias entirely by some delensing process; namely, $C_{\ell, \rm lens}^{BB} = N_\ell^{BB} = 0$.}
The $k_* = 7 \times 10^{-5} \, {\rm Mpc^{-1}}$ bispectrum is damped at $\ell \gtrsim 10$ (as shown in Fig.~\ref{fig:blllBBB}) and the SNR is saturated for $\ell_{\rm max} \gtrsim 10$. The $k_* = 5 \times 10^{-3} \, {\rm Mpc^{-1}}$ case is also undetectable, since the lensing bias is significantly large compared with the primary signal. We also find that, for the $k_* = 5 \times 10^{-4} \, {\rm Mpc^{-1}}$ case, the LiteBIRD noise does not reduce the SNR so much, since $N_{\ell, \rm LiteBIRD}^{BB}$ is as small as $C_{\ell, \rm lens}^{BB}$ at around the peak of the bispectrum $\ell \sim 50$, as shown in Fig.~\ref{fig:ClBB_pseudo}. The SNRs at $\ell_{\rm max} = 300$ for $k_* = 5 \times 10^{-4} \, {\rm Mpc^{-1}}$ are summarized in Table \ref{tab:SN_MFs}, indicating $4.6 \sigma$ and $2.5 \sigma$ detections in a noiseless full-sky survey (including the lensing bias) and LiteBIRD.

%%%%%%%%%%%%%%%%%%%%%%%%%%%%%%%%%%%%%%%%%%%%%%%%%%%%%%%%%%%%%%%%%%%%%

\section{B-mode Minkowski functional}\label{sec:MF}

The MF quantifies the topological information in CMB maps, such as area ($V_0$), circumference ($V_1$) and Euler characteristics ($V_2$). These quantities are sensitive to the statistics of CMB fluctuations, e.g., the skewness and the kurtosis, and hence one of useful NG indicators \cite{Mecke:1994ax,Schmalzing:1997aj,Matsubara:2003yt,Matsubara:2010te}. We here study the detectability of B-mode NG in the pseudoscalar model, by applying the analysis method for temperature and E-mode MFs \cite{Hikage:2008gy,Hikage:2012bs,Ade:2013ydc,Ade:2015ava} to the B-mode one. We then focus on only the $k_* = 5 \times 10^{-4} \, {\rm Mpc^{-1}}$ case, which is, among our choice of parameters, the sole case realizing a detectable B-mode NG, as seen in Sec.~\ref{sec:bis}. 

Assuming the weakness of B-mode NG, the MFs are perturbatively expressed as (up to $\sigma_0^2$)
\begin{eqnarray}
  V_k(\nu) = V_k^{({\rm G})}(\nu) + A_k e^{-\nu^2 / 2}
  \left[ v_k^{(1)}(\nu) \sigma_0 + v_k^{(2)}(\nu) \sigma_0^2  \right] ~, \label{eq:MF}
\end{eqnarray}
where $V_k^{({\rm G})}(\nu) \equiv A_k e^{-\nu^2 / 2} H_{k-1}(\nu)$ is the Gaussian part, $H_k(\nu)$ is the Hermite polynomials, and the amplitude is given by $A_k = (2\pi)^{-(k+1)/2} \frac{\omega_2}{\omega_{2 - k} \omega_k} ( \frac{\sigma_1}{\sqrt{2} \sigma_0} )^k$ with $ \omega_k \equiv \pi^{k/2} / \Gamma( \frac{ k}{2} + 1)$. This is the function of $\nu$, denoting a threshold value of the B-mode anisotropy ($B$) normalized by its standard deviation ($\sigma_0 \equiv \sqrt{\Braket{B^2}}$). The MFs are computed in the area of maps satisfying $B / \sigma_0 \geq \nu$. The harmonic-space representation of $\sigma_0$ and its derivative $\sigma_1 \equiv \sqrt{\Braket{|\nabla B|^2}}$ reads
\begin{eqnarray}
  \sigma_j^2 \equiv \frac{1}{4\pi} \sum_\ell (2\ell + 1) [\ell(\ell+1)]^j C_\ell^{BB} W_\ell^2 ~.
\end{eqnarray}
The window function $W_\ell$ filters the signal up to a threshold multipole $\ell_{\rm MF}$, taking $1$ for $\ell \leq \ell_{\rm MF}$ or $0$ for $\ell > \ell_{\rm MF}$.%
\footnote{If one chooses a Gaussian window smoothing over a threshold angular scale $\theta$, $W_\ell = \exp[-\ell(\ell+1) \theta^2 / 2]$, one can recover our results by taking $\theta \sim \pi / \ell_{\rm MF}$.}

In Eq.~\eqref{eq:MF}, the terms proportional to $\sigma_0$ and $\sigma_0^2$ contain NG information. The 1st-order term is the function of the skewness parameters, $S \equiv \Braket{B^3}/\sigma_0^4$, $S_{\rm I} \equiv B^2 \Braket{\nabla^2 B} / (\sigma_0\sigma_1)^2$ and $S_{\rm II} \equiv 2 \Braket{|\nabla B|^2 \nabla^2 B}/\sigma_1^4$, reading
\begin{eqnarray}
  v_k^{(1)}(\nu) = \frac{S}{6} H_{k+2}(\nu) - \frac{S_{\rm I}}{2} H_k(\nu) - \frac{S_{\rm II}}{2} H_{k-2}(\nu) ~. 
\end{eqnarray}
These skewness parameters are expressed in harmonic space as
\begin{eqnarray}
  S_A = \frac{1}{4\pi \sigma_0^4} \sum_{\ell_1 \ell_2 \ell_3} h_{\ell_1 \ell_2 \ell_3}
  \tilde{S}_{\ell_1 \ell_2 \ell_3}^A B_{\ell_1 \ell_2 \ell_3}^{BBB} W_{\ell_1} W_{\ell_2} W_{\ell_3} ~, \label{eq:skewparam}
  \end{eqnarray}
where $B_{\ell_1 \ell_2 \ell_3}^{BBB}$ is the angle-averaged B-mode bispectrum defined in Eq.~\eqref{eq:BBB_def}, and 
\begin{eqnarray}
  \tilde{S}_{\ell_1 \ell_2 \ell_3} &=& 1 ~, \\
  %----
  \tilde{S}_{\ell_1 \ell_2 \ell_3}^{\rm I}
  &=& - \frac{\sigma_0^2}{3\sigma_1^2}
\left( \{\ell_1 \} + \{\ell_2 \} + \{\ell_3 \} \right)
  ~, \\
%---
\tilde{S}_{\ell_1 \ell_2 \ell_3}^{\rm II}
  &=& \frac{\sigma_0^4}{3 \sigma_1^4}
\left( \{\ell_1 \}^2 + \{\ell_2 \}^2 + \{\ell_3 \}^2
  - 2 \{\ell_1\}\{\ell_2\} -2 \{\ell_2\}\{\ell_3\} - 2 \{\ell_3\}\{\ell_1\} \right)
~,
  \end{eqnarray}
with $\{\ell \} \equiv \ell(\ell+1)$. The skewness equates to the
angle-averaged quantity of the three-point correlation of the fields
at the identical point $\hat{n}$. The B-mode field and its derivatives
are spin-0 and therefore their angle dependences are completely
characterized by $Y_{\ell m}(\hat{n})$ (under an assumption of the flat universe). Averaging over all directions of $\hat{n}$ then results in
the so-called Gaunt integral $\int d^2 \hat{n} Y_{\ell_1 m_1}(\hat{n})
Y_{\ell_2 m_2}(\hat{n}) Y_{\ell_3 m_3}(\hat{n})$, leading to
$h_{\ell_1 \ell_2 \ell_3}$ in Eq.~\eqref{eq:skewparam}. The
geometrical factor $h_{\ell_1 \ell_2 \ell_3}$ filters only the signals
satisfying $\ell_1 + \ell_2 + \ell_3 = {\rm even}$, meaning that the MFs
do not include any $\ell_1 + \ell_2 + \ell_3 = {\rm odd}$ information
of the B-mode auto-bispectrum. This is advantageous to our
analysis. The $\ell_1 + \ell_2 + \ell_3 = {\rm odd}$ components in the
BBB bispectrum are contaminated by the lensed signals due to the
scalar mode (see Appendix~\ref{appen:bis_lens}), while we do not need to worry about this lensing bias in the MF analysis.

MFs also contain the 2nd-order term $v_k^{(2)}$, depending on the
kurtosis as well as the skewness \cite{Matsubara:2010te}. This term is
highly contaminated with the lensing signal via the kurtosis
\cite{Zhao:2015sla}. This 2nd-order term is mathematically orthogonal to the 1st-order term and each information can be independently measured in the MF analysis. To extract the clean information on the primary B-mode NG, we deal with only $v_k^{(1)}$ and discard the information from $v_k^{(2)}$ in our analysis.

\begin{figure}[t!]
  \begin{tabular}{c}
    \begin{minipage}{1.0\hsize}
  \begin{center}
    \includegraphics[width = 0.85\textwidth]{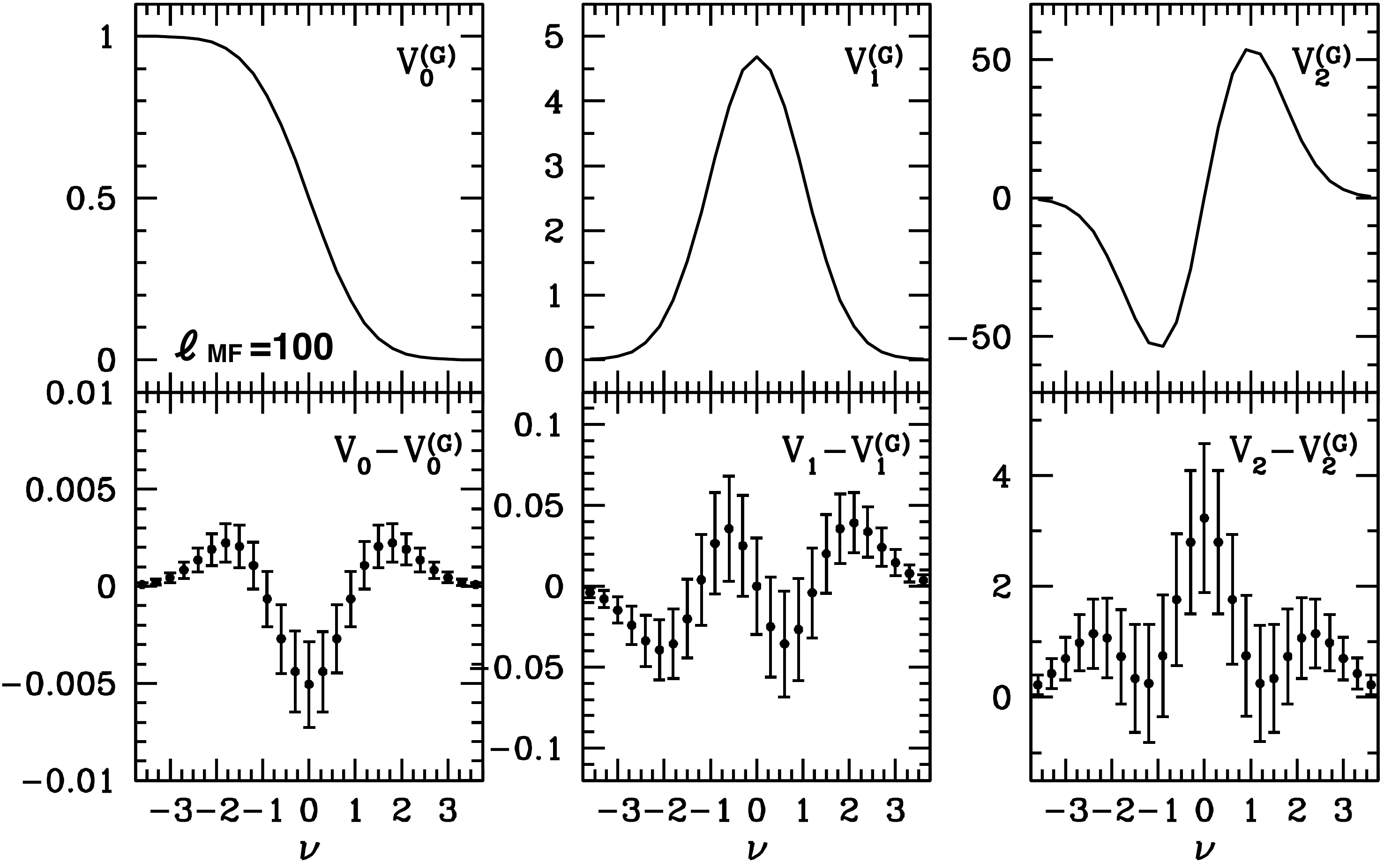}
  \end{center}
\end{minipage}
\end{tabular}
  \caption{Minkowski functionals for B-mode maps with $\ell_{\rm
      MF}=100$, originating from the pseudoscalar model with $k_* = 5 \times 10^{-4} \, {\rm Mpc^{-1}}$ (upper panel) and their residuals from the Gaussian MFs (lower panel).  Error bars represent the $1\sigma$ dispersion
    estimated from 10000 all-sky Gaussian realizations with
    $C_{\ell, \rm dat}^{BB}=C_{\ell, \rm prim}^{BB}$.}\label{fig:MFs}
\end{figure}

Figure~\ref{fig:MFs} shows the MFs and their differences from
the Gaussian contributions for B-mode maps with $\ell_{\rm MF}=100$, expected in the pseudoscalar model with $k_* = 5 \times 10^{ -4} \, {\rm Mpc^{-1}}$.  The
B-mode power spectrum is set to be the primordial one, i.e.,
$C_{\ell, \rm dat}^{BB}=C_{\ell, \rm prim}^{BB}$ and lensed B-mode is not included in these expectations.%
\footnote{Even if the lensing is included, the results would not be significantly affected, as we include only up to $\ell_{\rm MF} = 100$.}
The errors are numerically estimated from Gaussian realizations. From this figure, at $\nu \sim 0$ in $V_{0,2}$, more than $1 \sigma$ deviation from Gaussianity is clearly seen.

The SNR in the MF analysis is expressed as
\begin{eqnarray}
  \left( \frac{S}{N} \right)_{\rm MF}^2 = \sum_{ij}
  \left(V_i-V_i^{\rm (G)}\right) \left({\rm Cov}^{-1}\right)_{ij}
  \left(V_j-V_j^{\rm (G)}\right)
  \end{eqnarray}
where the summation is over different bins of $\nu$, different
$\ell_{\rm MF}$ and three kinds of the MFs. The binning number of $\nu$
is set to be 25 from $-3.75$ to $3.75$. We use the MFs for B-mode maps
with different $\ell_{\rm MF}$ (30, 50, 70, 100, and 200) to extract the information of scale dependence. We checked that 5 bins are enough to obtain the accurate value of the SNR after summing up all MFs. The MFs are then correlated among different $\nu$, $k$, and $\ell_{\rm MF}$. We numerically evaluate the covariance using 10000 Gaussian realizations, which is enough to converge our estimates. As we already see in the previous section that the SNR is saturated at $\ell_{\rm max} = 200$, we restrict our MF analysis to $\ell_{\rm MF} \leq 200$ and omit any NG contributions (which are effective only for higher $\ell$) from the covariance matrix.

Table \ref{tab:SN_MFs} lists the SNRs of the B-mode NG from the MFs for B-mode maps with different $\ell_{\rm MF}$ values and with all of them combined. We take into account the three measurements: a noiseless all-sky survey without the lensing bias ($C_{\ell, \rm dat}^{BB} = C_{\ell, \rm prim}^{BB}$ and $f_{\rm sky} = 1$), a noiseless all-sky survey with the lensing bias ($C_{\ell, \rm dat}^{BB} = C_{\ell, \rm prim}^{BB} + C_{\ell, \rm lens}^{BB}$ and $f_{\rm sky} = 1$), and LiteBIRD ($C_{\ell, \rm dat}^{BB} = C_{\ell, \rm prim}^{BB} + C_{\ell, \rm lens}^{BB} + N_{\ell, \rm LiteBIRD}^{BB}$ and $f_{\rm sky} = 0.5$). We find that, for the cases including nonzero $C_{\ell, \rm lens}^{BB}$, the SNRs of the MFs are highest at $\ell_{\rm MF} \sim 70$, which is consistent with the scale where the SNRs of B-mode bispectra converge (see Fig.~\ref{fig:SNBBB_even}). As usual, the SNRs from combined MFs reach roughly half compared with those from the B-mode bispectra. Unfortunately, there should be no $2\sigma$ detection in LiteBIRD, but a noiseless survey should catch $3 \sigma$ signal even without delensing. We also find that in an ultimate ideal case where $C_{\ell, \rm lens}^{BB} = N_\ell^{BB} = 0$, the SNR can potentially improved up to $6.4$.

\begin{table}[t]
 \begin{center}
   \begin{tabular}[c]{|ccc|c||cccccc|c|}\hline
 \multicolumn{3}{|c|}{$C_{\ell,\rm dat}^{BB}$} & \multirow{2}{*}{$f_{\rm sky}$} & \multicolumn{6}{c|}{$(S/N)_{\rm MF}$} & \multirow{2}{*}{$(S/N)_{BBB}$} \\ %\cline{4-9}
 $C_{\ell, \rm prim}^{BB}$ & $C_{\ell, \rm lens}^{BB}$ & $N_{\ell}^{BB}$ & & 30 & 50 & 70 & 100 & 200 & total &  \\ \hline\hline
 $\checkmark$ & &  & 1 & 1.6 & 2.8 & 3.5 & 4.4 & 5.7 & 6.4 & 10  \\ 
 $\checkmark$ & $\checkmark$ & & 1 & 1.2 & 2.1  & 2.3 & 2.2 & 1.7 & 3.0 & 4.6 \\
 $\checkmark$ & $\checkmark$ & $\checkmark$ & 0.5 & 0.66 & 1.1 & 1.2 & 1.0 & 0.78 & 1.6 & 2.5 \\ \hline
 \end{tabular}
 \end{center}
 \caption{SNRs of the B-mode NG in the pseudoscalar model with $k_* = 5 \times 10^{-4} \, {\rm Mpc^{-1}}$ obtained from MFs for B-mode maps for each $\ell_{\rm MF}$ (30, 50, 70, 100 and 200) and the total SNR summed over all $\ell_{\rm MF}$ bins. We here consider the measurements with three different $(C_{\ell, \rm dat}^{BB}, f_{\rm sky})$. The absence of $C_{\ell, \rm prim}^{BB}$, $C_{\ell, \rm lens}^{BB}$ or $N_{\ell}^{BB}$ corresponds to null-hypothesis tests, measurements with a perfect delensing, or noise-free measurements. For comparison, in the rightmost column, we show the SNRs computed in the harmonic-space bispectrum analysis, where the result for $C_{\ell, \rm dat}^{BB}=C_{\ell, \rm prim}^{BB}$ is obtained in \cite{Namba:2015gja}, while the others are obtained for the first time in this paper.}\label{tab:SN_MFs}
 \end{table}

%%%%%%%%%%%%%%%%%%%%%%%%%%%%%%%%%%%%%%%%%%%%%%%%%%%%%%%%%%%%%%%%%%%%%
\section{Vacuum or source?}\label{sec:comparison}

\begin{figure}[t!]
  \begin{tabular}{c}
    \begin{minipage}{1.0\hsize}
  \begin{center}
    \includegraphics[width = 0.85\textwidth]{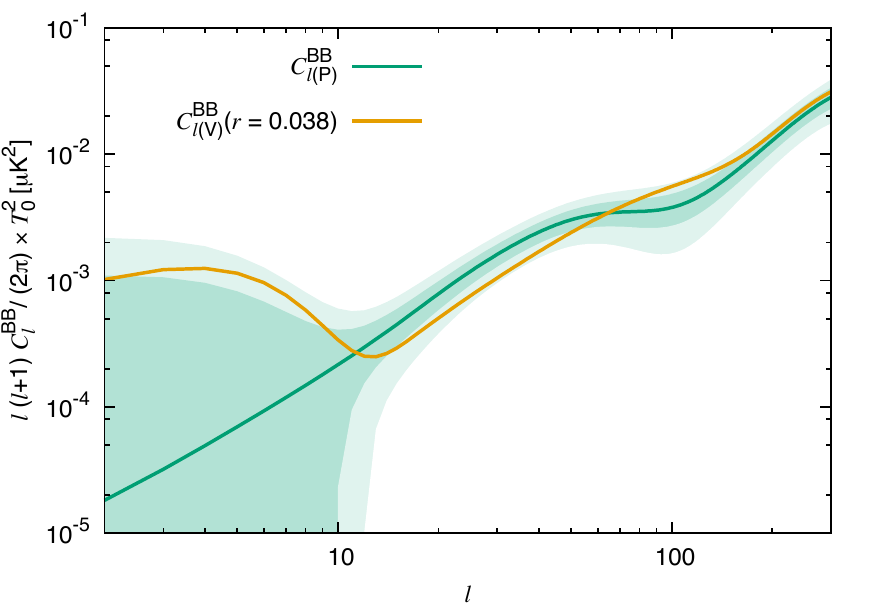}
  \end{center}
\end{minipage}
\end{tabular}
  \caption{Vacuum-mode BB spectrum minimizing $\chi_{BB}^2$ \eqref{eq:chisq_BB} with $N_\ell^{BB} = N_{\ell, \rm LiteBIRD}^{BB}$, $f_{\rm sky} = 0.5$, $\ell_{\rm min} = 2$ and $\ell_{\rm max} = 300$, i.e., $C_{\ell \rm (V)}^{BB}(r = r_{\rm bf} = 0.038)$ (yellow line) and the fitted BB spectrum, i.e., $C_{\ell \rm (P)}^{BB}$ (green solid line). Dark (pale) green region describes $<1\sigma$ ($< 2 \sigma$) uncertainties. 
It is clearly seen that the vacuum-originated spectrum $C_{\ell \rm (V)}^{BB}(r = 0.038)$ is completely within $2\sigma$ (pale green region) of the fiducial (pseudoscalar) spectrum,
 as expected from $\chi_{{\rm red}, BB}^2(r = 0.038) = 1.1$ (see Table~\ref{tab:chisq}).}\label{fig:ClBB_vac_bestfit_LiteBIRD} 
\end{figure}

From the results in Secs.~\ref{sec:bis} and \ref{sec:MF} and Appendix~\ref{appen:pow_BB}, we understand that, if the pseudoscalar model with $k_* = 5 \times 10^{-4} \, {\rm Mpc}^{-1}$ were the true model, the BBB bispectrum and/or the B-mode MF, in addition to the BB power spectrum, would also be detected, owing to the existence of strong NG source.
In this case, however, a key question would be whether we can really confirm this model from the data and unambiguously exclude the alternatives. As an example, we here assume a standard (quasi) scale-invariant B-mode spectrum induced by the parity-conserving GWs originating from vacuum fluctuations, against which the true model should be distinguished. On the other hand, one may as well postulate scale dependence in analyzing (future) real data. However, if such an additional parameter is introduced, the fitting of the data would in general become better. In this sense, our assumption of a scale-independent power spectrum of the vacuum GW mode is a conservative one. As demonstrated below, even in this conservative case the BB power spectrum in our model is not completely distinguished from the standard vacuum GWs. The crucial message here is that the complete discrimination between vacuum and sourced modes should thus resort to the analysis of BBB bispectrum.

We investigate the capability of upcoming, LiteBIRD-like experiments to discriminate a sourced B-mode against a vacuum one, by evaluating the deviation of the latter from the data. To this end, we assume the true model to be the pseudoscalar one we study in this paper and the observed spectra to be originating from it with $k_* = 5 \times 10^{-4} \, {\rm Mpc}^{-1}$, corresponding to the green curves in Fig.~\ref{fig:blllBBB}. We first examine goodness of fit for the BB power spectrum by computing the reduced $\chi^2$ of the best-fit model. We define the standard $\chi^2$ for BB as
  \begin{eqnarray}
    \chi_{BB}^2(r) = \sum_{\ell = \ell_{\rm min}}^{\ell_{\rm max}} \frac{2\ell + 1}{2} \left(\frac{C_{\ell ({\rm V})}^{BB}(r)  - C_{\ell ({\rm P})}^{BB} }{C_{\ell ({\rm V})}^{BB}(r)  + N_\ell^{BB}} \right)^2 ~, \label{eq:chisq_BB}
  \end{eqnarray}
  where we refer to the vacuum-mode spectrum (parametrized by the usual tensor-to-scalar ratio $r$) and that in the pseudoscalar model as (V) and (P), respectively, and $C_\ell^{BB}$ includes both primary and lensed BB contributions. The best-fit model is given by $r$ minimizing $\chi_{BB}^2$, referred to as $r_{\rm bf}$. 
  We further define the reduced $\chi^2$ as $\chi_{{\rm red,} BB}^2 \equiv \chi_{BB}^2 / ( \ell_{\rm max} - \ell_{\rm min} )$, with $r$ the only fitted parameter. In the case that the reduced $\chi^2$ with $r_{\rm bf}$ given is not much larger than unity, the vacuum-mode spectrum and the pseudoscalar one are indistinguishable from each other. Further discriminations with the NG observables as the BBB bispectrum and the MF then become important. The $\chi^2$ for the BBB bispectrum is given as
\begin{eqnarray}
    \chi_{BBB}^2(r) = \sum_{\ell_1, \ell_2, \ell_3 = \ell_{\rm min}}^{\ell_{\rm max}}
    \frac{\left| B_{\ell_1 \ell_2 \ell_3 ({\rm P})}^{BBB}\right|^2}{6 \prod_{n=1}^3 \left(C_{\ell_n ({\rm V})}^{BB}(r) + N_{\ell_n}^{BB} \right)} ~.
\label{chi2BBB}
\end{eqnarray}
Here we limit the summation range to $\ell_1 + \ell_2 + \ell_3 = {\rm even}$ as usual in this paper and hence there is no vacuum-mode contribution in the numerator.

We compute the $\chi^2$ values of the B-mode power spectrum, bispectrum and MF for the pseudoscalar model by assuming a LiteBIRD-like survey. Our results are summarized in Table~\ref{tab:chisq}. 
In a realistic setup with the noise spectrum $N_\ell^{BB}$ included and with $f_{\rm sky} = 0.5$, we obtain $\chi_{{\rm red,} BB}^2 (r_{\rm bf}) \simeq 1$, implying that the BB analysis can provide the fitting such that the vacuum and pseudoscalar B-mode power spectra are compatible within $\sim 1 \sigma$ at each bin. This is also visually confirmed in Fig.~\ref{fig:ClBB_vac_bestfit_LiteBIRD}, where the vacuum spectrum (yellow curve) is near the edge of the $1\sigma$ region (dark green) of the pseudoscalar spectrum (green curve) and completely inside the $2\sigma$ region (pale green). In such cases as this, it is necessary to push the analysis to bispectra in order to distinguish the two different models. Indeed, the large values of $\chi_{BBB}^2(r_{\rm bf})$ and $\chi_{\rm MF}^2(r_{\rm bf})$ in Table~\ref{tab:chisq} are quite promising. Namely, the BBB (MF) analysis with the LiteBIRD data could exclude the models dominated by vacuum fluctuations, with a $3.6\sigma$ ($2.2 \sigma$) significance. This gives us a strong motivation to search for B-mode NGs.

We also notice that such high statistical significances are supported by 
the lowest value of $\ell$ we can use for the analysis, denoted as $\ell_{\rm min}$. For comparison, we take $\ell_{\rm min} = 2$ and $31$ in Table~\ref{tab:chisq} and find that they fall below $2\sigma$, as $\ell_{\rm min}$ increases. This suggests the particular need for space-based measurements such as LiteBIRD in order to achieve a high performance for model discriminability, since such low $\ell$ regimes are out of reach of ground- or balloon-based observations.
Table~\ref{tab:chisq} also contains the results of an ideal noiseless all-sky survey, showing that, even without delensing, the reduction of instrumental noise level enables more accurate model discrimination.

\begin{table}[t]
 \begin{center}
   \begin{tabular}[c]{|c|c|c|c||cccccc|c|}\hline
     \multirow{2}{*}{$N_{\ell}^{BB}$} & \multirow{2}{*}{$f_{\rm sky}$} & \multirow{2}{*}{$\chi_{{\rm red}, BB}^{2}(r_{\rm bf})$} & \multirow{2}{*}{$r_{\rm bf} \times 10^2$} & \multicolumn{6}{c|}{$\chi_{\rm MF}^2(r_{\rm bf})$} & \multirow{2}{*}{$\chi_{BBB}^2(r_{\rm bf})$} \\ %\cline{4-9}
     & & & & 30 & 50 & 70 & 100 & 200 & total &  \\ \hline\hline
 & 1 & 4.8 (4.6) & 3.7 (3.4) & 4.6 & 9.9 & 6.1 & 2.0 & 0.4 & 19 & 54 (14) \\
  $\checkmark$ & 0.5 & 1.1 (1.0) & 3.8 (3.5) & 1.1 & 2.6 & 1.6 & 0.5 & 0.1 & 4.8 & 13 (3.5) \\ \hline
 \end{tabular}
 \end{center}
 \caption{The $\chi^2$ values for the B-mode MFs (fifth line-separated column) and the BBB bispectra (sixth column) generated from the vacuum-mode spectra with $r = r_{\rm bf}$, under the assumption that the pseudoscalar model with $k_* = 5 \times 10^{-4} \, {\rm Mpc^{-1}}$ is the true model. We here consider a noiseless full-sky survey (first row) and a LiteBIRD-like one (second row). The minimum reduced $\chi^2$ and the corresponding best-fit $r$, estimated from the BB power spectra with $\ell_{\rm max} = 300$, are described in the third and fourth columns, respectively. The settings in the MF analysis are the same as in Table~\ref{tab:SN_MFs}. The results in the brackets are obtained with $\ell_{\rm min} = 31$, but all others are based on $\ell_{\rm min} = 2$.
}\label{tab:chisq}
 \end{table}

%%%%%%%%%%%%%%%%%%%%%%%%%%%%%%%%%%%%%%%%%%%%%%%%%%%%%%%%%%%%%%%%%%%%%
\section{Conclusions}\label{sec:conclusions}

Once primordial GWs are detected in a future survey of CMB B-mode polarization, the next critical step would be to seek its origin, namely to judge whether the B-mode arises from vacuum fluctuations during inflation or some other sources. Testing statistics of observed B-mode fluctuations will give an important clue for answering this question. The vacuum mode is expected to be nearly Gaussian, but the mode sourced by other fields could have large NG. Since the NG is induced by interactions during the time of GW production, its specific features are subject to specific mechanisms. Most of the inflationary models do not produce large GW NG and hence large B-mode NG within the parameter space that satisfies observational constraints from the current CMB temperature and polarization data. There are however exceptional cases that need to be studied extensively.

To demonstrate the need for NG tests of the B-mode polarization, we have considered a concrete model in which a $U(1)$ gauge field produced through a coupling to a pseudoscalar gravitationally sources parity-violating NG GWs. We have examined the detectability of the induced B-mode NG and the distinguishability between the GWs in this model and the standard vacuum-induced ones, assuming a proposed full-sky survey satellite LiteBIRD. We have employed two well-known NG estimators, the harmonic-space bispectrum estimator and the MF estimator. We have restricted our analysis to the $\ell_1 + \ell_2 + \ell_3 = {\rm even}$ modes, which stay non-vanishing only when three-point correlation functions of GWs have unequal amplitudes among correlations of different circularly-polarized states. In this way, we are able to estimate the primordial parity-violating B-mode NG without a bias from the scalar-mode lensing contribution. Interestingly, the B-mode MF is independent of the $\ell_1 + \ell_2 + \ell_3 = {\rm odd}$ modes, due to parity-conserving nature of the skewness parameters. It is hence a clean observable of parity-violating B-mode bispectrum, which we have confirmed in this study.
 
In our forecasts, we have taken into account specific instrumental features (beam sizes, the sky coverage and instrumental noise) expected for the LiteBIRD mission. Uncertainties due to residual galactic foregrounds and late-time lensing signals also been estimated. These affect the covariance matrix of the bispectrum or the MF, and reduce the sensitivity. Nevertheless, we have found that the GW NG can be measured with a SNR of $2.5$ by employing the harmonic-space bispectrum estimator. The analysis of BB power spectrum alone is not necessarily able to distinguish the scale-dependent parity-violating GWs predicted in the model of our interest from the usual vacuum-induced ones because of the shape similarity within the range of cosmic variance, as seen in Fig.~\ref{fig:ClBB_vac_bestfit_LiteBIRD}. The signal of the BBB bispectrum can, on the other hand, allow us to rule out the case with vacuum fluctuation only with a $3.6 \sigma$ significance with the LiteBIRD data for the model we consider in this work. While the sensitivity achieved in the MF estimator is somewhat lower than that obtained in the bispectrum estimator, the former is still useful for validation of the results obtained by the latter. The series of BB and BBB analyses we have described convey an important message; provided a B-mode detection, exploration of its NG signatures, in particular the BBB signal, will be indispensable toward unambiguously establishing the quantum fluctuation of vacuum as the origin of the GW.

The model considered in this paper also produces nonzero TB and EB correlations with specific scale dependence. Including uncertainties due to lensing and the noise spectrum, $\text{SNR} \simeq 1.2$ is obtained from a joint TB+EB analysis,%
\footnote{Here, we simply assume that the data of the EB correlation can be used to measure the chiral GW, although LiteBIRD plans to use this information for calibration of the polarization angle.}
in a LiteBIRD-like measurement. Although this bare value is not competitive, a combined analysis of power spectra and bispectra in principle enhances the statistical significance. It thus helps constrain the model and clarify the presence or absence of source fields producing chiral GWs.
 
Besides the above known observables, we have found new parity-violating observables that were not investigated previously. Similar to the B-mode polarization, the curl mode of the gravitational lensing is generated by the vector and tensor perturbations and has odd-parity symmetry \cite{Dodelson:2003bv,Cooray:2005hm,Li:2006si,Namikawa:2011cs}. In the presence of the curl mode ($\omega$), there are correlations between the E-mode polarization and $\omega$ ($E\omega$), and the lensing potential and $\omega$ ($\phi\omega$). In addition, the lensing produces the correlation between the B-mode polarization and the lensing potential ($B\phi$), and the $\ell_1 + \ell_2 + \ell_3 = {\rm even}$ components of the BBB bispectrum. While these new observables are undetectably small in the model with the parameter region we have focused on, they could potentially provide some clues on the physics behind the GW production in the early universe, if they could be at a detectable level through some enhancement mechanism.

%%%%%%%%%%%%%%%%%%%%%%%%%%%%%%%%%%%%%%%%%%%%%%%%%%%%%%%%%
\acknowledgments

We thank Yoshihiko Oyama, Marco Peloso, Shohei Saga, Lorenzo Sorbo and Caner Unal for useful discussions. MS was supported in part by a Grant-in-Aid for JSPS Research under Grants No.~27-10917. TN is grateful for a support from the JSPS Postdoctoral Fellowships for Research Abroad (No.~26-142). MH was supported by MEXT KAKENHI Grant Number 15H05891. MS, CH, RN and MH were supported in part by the World Premier International Research Center Initiative (WPI Initiative), MEXT, Japan. Numerical computations by MS were in part carried out on Cray XC30 at Center for Computational Astrophysics, National Astronomical Observatory of Japan.

%%%%%%%%%%%%%%%%%%%%%%%%%%%%%%%%%%%%%%%%%%%%%%%%%%%%%%%%%%%%%%%%%

\appendix
%%%%%%%%%%%%%%%%%%%%%%%%%%%%%%%%%%%%%%%%%%%%%%%%%%%%%%%%%

\section{Power spectrum analysis}\label{appen:pow_BB}

\begin{figure}[t!]
  \begin{tabular}{c}
    \begin{minipage}{1.0\hsize}
  \begin{center}
    \includegraphics[width = 0.85\textwidth]{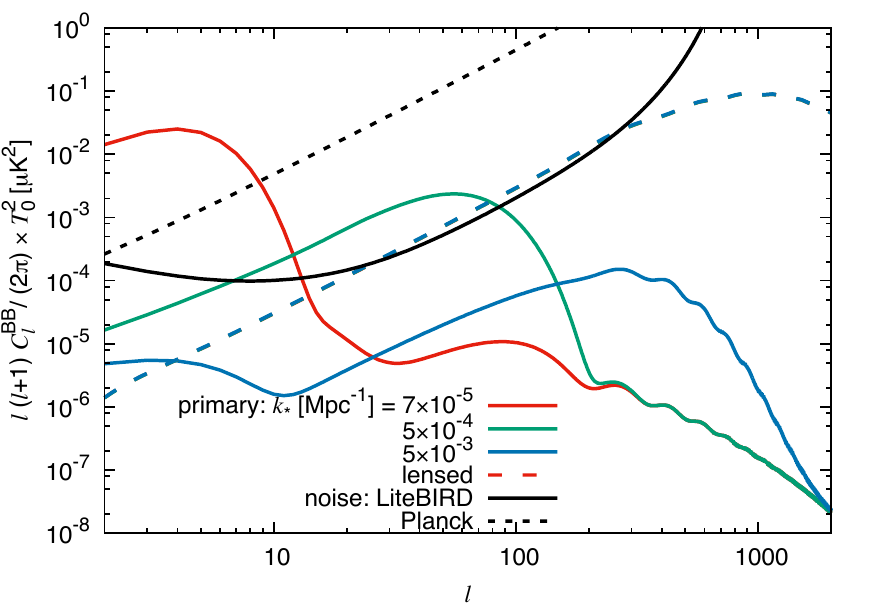}
  \end{center}
\end{minipage}
\end{tabular}
  \caption{Primary B-mode power spectra $C_{\ell, \rm prim}^{BB}$ (colored solid lines) and lensed B-mode ones $C_{\ell, \rm lens}^{BB}$  (colored dashed lines) for three different $k_*$ in the pseudoscalar model \cite{Namba:2015gja}. For comparison, the noise spectra (including residual foreground) in LiteBIRD \eqref{eq:NlPP_LiteBIRD} (black solid line) and {\it Planck} \cite{Planck:2006aa} (black dotted line) are also plotted. Note that all colored dashed lines almost overlap each other.}\label{fig:ClBB_pseudo} 
\end{figure}

\begin{figure}[t]
  \begin{tabular}{cc}
    \begin{minipage}{0.5\hsize}
  \begin{center}
    \includegraphics[width=1\textwidth]{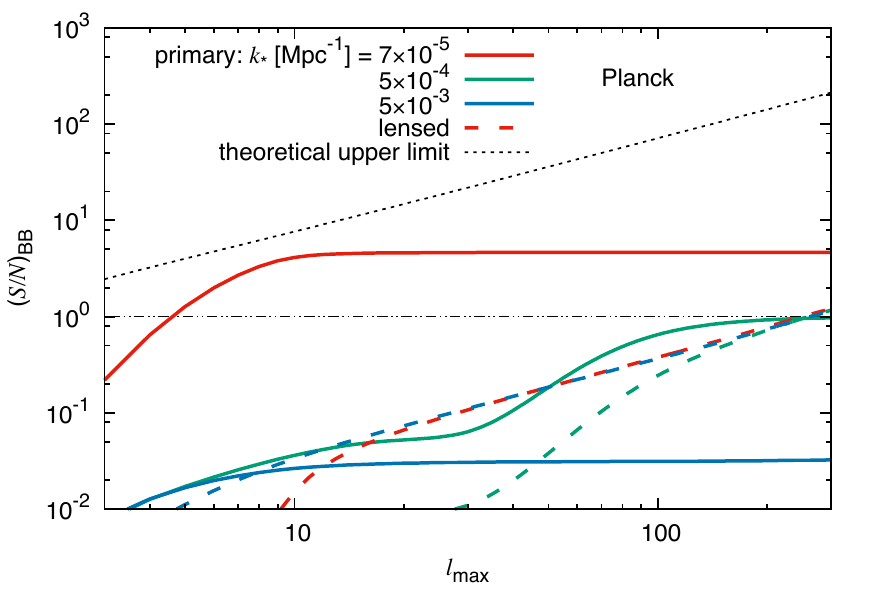}
  \end{center}
\end{minipage}
\begin{minipage}{0.5\hsize}
  \begin{center}
    \includegraphics[width=1\textwidth]{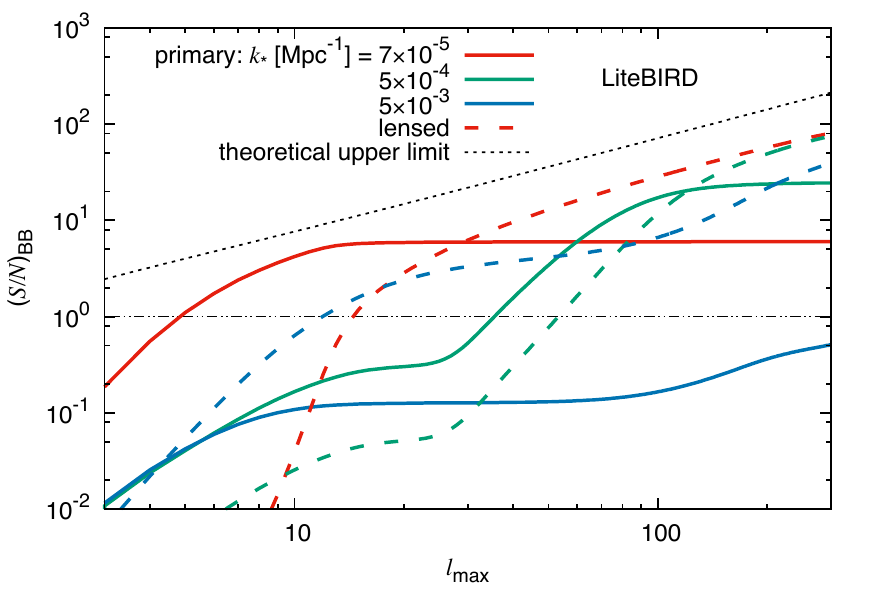}
  \end{center}
\end{minipage}
\end{tabular}
  \\
    \begin{tabular}{c}
    \begin{minipage}{1.0\hsize}
  \begin{center}
    \includegraphics[width = 0.5\textwidth]{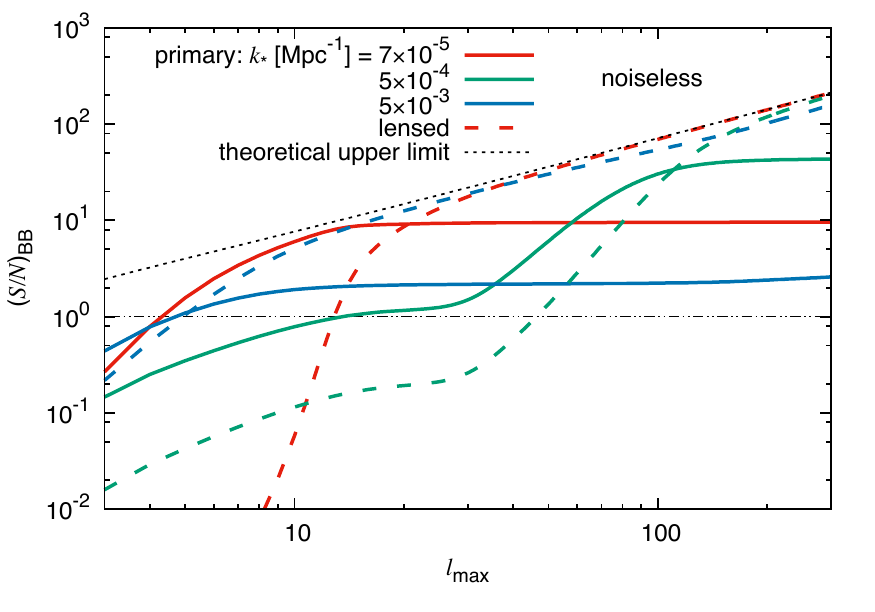}
  \end{center}
\end{minipage}
\end{tabular}
    \caption{SNRs of the primary B-mode power spectra $(S/N)_{\rm prim}$ (solid lines) and the lensed B-mode ones $(S/N)_{\rm lens}$ (dashed lines) as functions of $\ell_{\rm max}$, expected in a {\it Planck}-like ($f_{\rm sky} = 0.7$), LiteBIRD-like ($f_{\rm sky} = 0.5$) and noiseless ($f_{\rm sky} = 1$) survey. For each case, the primary and lensed signals are estimated jointly. We here consider the cases for three different $k_*$ in the pseudoscalar model. The black dotted lines show the theoretical upper limits; namely, $(S/N)^2 = \sum_{\ell=2}^{\ell_{\rm max}}(2\ell + 1)/2 = (\ell_{\rm max}+3)(\ell_{\rm max}-1)/2$.}
\label{fig:SNBB}
\end{figure}

Here, we compute SNRs of the primary BB power spectra in the pseudoscalar model, assuming a LiteBIRD-like survey. The primary power spectra, peaking at $\ell \sim k_* \tau_0$, are shown in Fig.~\ref{fig:ClBB_pseudo} and compared with the lensed BB power spectrum and the noise power spectra in ${\it Planck}$ \cite{Planck:2006aa} and LiteBIRD~\eqref{eq:NlPP_LiteBIRD}.

Due to gravitational lensing at late times, primary scalar-mode E-mode fluctuations are partially converted into B-mode fluctuations. To know SNRs of the primary tensor-mode signal under the presence of the lensing bias, we compute the 2D Fisher matrix
\begin{eqnarray}
  F_{ij} \equiv \sum_{\ell = 2}^{\ell_{\rm max}} \frac{2\ell + 1}{2} \frac{C_{\ell, i}^{BB} C_{\ell, j}^{BB}}{(C_{\ell,\rm dat}^{BB})^2} ~,
\end{eqnarray}
and estimate
\begin{eqnarray}
  \left( \frac{S}{N} \right)_{i}^2 = \frac{1}{(F^{-1})_{ii}} ~,
  \end{eqnarray}
where $i,j$ run over ``prim'' and ``lens''. Here we have ignored any NG contributions due to lensing in the covariance matrix, since they are very weak (at most sub-percent level) on our interesting scales $\ell \leq 300$ \cite{BenoitLevy:2012va,Namikawa:2015tba}. 

The $\ell_{\rm max}$ dependences of $(S/N)_{\rm prim}$ and $(S/N)_{\rm lens}$ in a {\it Planck}-like, LiteBIRD-like and ideal noiseless experiments are shown in Fig.~\ref{fig:SNBB}. It is obvious in this figure that for the LiteBIRD and noiseless case, $(S/N)_{\rm prim}$ defeat $(S/N)_{\rm lens}$ for $\ell_{\rm max} \lesssim 20 $ (when $k_* = 7 \times 10^{-5} \, {\rm Mpc^{-1}}$), $\ell_{\rm max} \lesssim 100 $ (when $k_* = 5 \times 10^{-4} \, {\rm Mpc^{-1}}$) and $\ell_{\rm max} \lesssim 5 $ (when $k_* = 5 \times 10^{-3} \, {\rm Mpc^{-1}}$). This matches the expectations from the magnitude relations between $C_{\ell, \rm prim}^{BB}$ and $C_{\ell, \rm lens}^{BB}$ described in Fig.~\ref{fig:ClBB_pseudo}. Fig.~\ref{fig:SNBB} also shows that {\it Planck} is hard to observe the primary signal, except in the $k_* = 7 \times 10^{-5} \, {\rm Mpc^{-1}}$ case, due to sizable instrumental noise. Note that $(S/N)_{\rm prim}$ in {\it Planck} is a bit smaller than the results in \cite{Namba:2015gja}, since we here take into account the contamination by lensed B-mode and the loss of information due to a 70\% sky coverage.

%%%%%%%%%%%%%%%%%%%%%%%%%%%%%%%%%%%%%%%%%%%%%%%%%%%%%%%%%%%%%%%%%%%%%
\section{Lensed B-mode bispectrum}\label{appen:bis_lens}

In this Appendix, we discuss the effect of lensing on the BBB bispectrum.
Ignoring the curl mode, the B-mode modulated by lensing is given by \cite{Hu:2000ee,Lewis:2011fk,Pearson:2012ba}
\begin{eqnarray}
\tilde{a}_{\ell m}^B = a_{\ell m}^B + \sum_{LM} \sum_{l' m'}
  \left(
  \begin{array}{ccc}
  \ell & L & l' \\
  m & M & m' 
  \end{array}
\right)
\phi_{LM}^* \left[ 
F_{\ell L l'}^{+} a_{l' m'}^{B *} + i F_{\ell L l'}^{-} a_{l' m'}^{E *}
  \right] + {\cal O}(\phi^2 (a^{E} + a^B)) ~, 
\end{eqnarray}
where $a_{\ell m}^{E/B}$ is the unlensed primary E/B-mode anisotropy, $\phi_{LM}$ is the harmonic coefficients of the lensing potential, and 
\begin{eqnarray}
  F_{l_1 l_2 l_3}^{\pm} &\equiv& \frac{l_2(l_2 + 1) + l_3(l_3 + 1) - l_1 (l_1 + 1)}{2} h_{l_1 l_2 l_3 }^{2 0 -2}
  \frac{1 \pm (-1)^{l_1 + l_2 + l_3}}{2} ~.
\end{eqnarray}
Note that $F_{l_1 l_2 l_3}^{+/-}$ can take nonzero values only for $l_1 + l_2 + l_3 = {\rm even / odd}$, and $h_{l_1 l_2 l_3 }^{2 0 -2}$ is given by Eq.~\eqref{eq:hsym}. On large scales under examination, $\ell < 300$, we may treat $a_{\ell m}^{E/B}$ and $\phi_{\ell m}$ as Gaussian fields, since both primary and lensed NGs are expected to be very weak. The lensed bispectrum is then decomposed using the power spectra $C^{XY}$ by means of the Wick theorem and we obtain
\begin{eqnarray}
  B_{\rm lens}^{BBB}
  &\sim& F^+ C^{B \phi} C^{BB} + F^{-} C^{B \phi} C^{EB} 
   \nonumber \\ 
   %-----
   && + F^+ F^+ F^+ \left[ C^{B\phi}C^{B\phi}C^{B\phi} + C^{B\phi} C^{BB} C^{\phi \phi}  \right] \nonumber \\
   && + F^+ F^+ F^{-} \left[ C^{B\phi} C^{B\phi} C^{E\phi} + C^{B\phi} C^{BE} C^{\phi\phi} + C^{E\phi} C^{BB} C^{\phi\phi}   \right] \nonumber \\
   && + F^{+} F^- F^{-} \left[ C^{B\phi}C^{E\phi}C^{E\phi}
     + C^{B\phi} C^{EE} C^{\phi \phi} + C^{E\phi} C^{BE} C^{\phi \phi} \right] \nonumber \\ 
   && + F^- F^- F^- \left[ C^{E\phi}C^{E\phi}C^{E\phi} + C^{E\phi} C^{EE} C^{\phi \phi}  \right] + {\cal O}(\phi^5) ~,
\end{eqnarray}
where we omit prefactors and arguments to see the broad structure of $B_{\rm lens}^{BBB}$. The 1st line corresponds to the tree-level expression, while the terms on and after the 2nd line include the 1-loop computations. The combination of $F^+$ and $F^{-}$ determines parity; namely, the terms including even/odd number of $F^{-}$ limit nonvanishing signals to $\ell_1 + \ell_2 + \ell_3 = {\rm even/ odd}$.

In the absence of the primary GW, nonvanishing signal comes only from the terms including $C^{E\phi}C^{E\phi}C^{E\phi}$ and $C^{E\phi} C^{EE} C^{\phi \phi}$, which are limited to $\ell_1 + \ell_2 + \ell_3 ={\rm odd}$. In search for the primary GW, this scalar-mode contribution behaves as a bias, and thus we analyze only the $\ell_1 + \ell_2 + \ell_3 = {\rm even}$ domain in Secs.~\ref{sec:bis} and \ref{sec:MF}, which completely reduces the lensing contamination.

On the other hand, in the pseudoscalar model, there exist nonvanishing parity-violating correlators $C^{B\phi}$ and $C^{EB}$, inducing nonvanishing lensed BBB bispectrum in $\ell_1 + \ell_2 + \ell_3 = {\rm even}$. Focusing on this domain, we analyze this bispectrum with the combination of $F^+ C^{B \phi} C^{BB}$ , $F^+ F^- F^- C^{B \phi}C^{E\phi}C^{E\phi}$ and $F^+ F^- F^- C^{B \phi} C^{EE} C^{\phi \phi}$, and ignore the other subdominant terms.%
\footnote{
We here neglect the $F^+ F^- F^- C^{E \phi} C^{BE} C^{\phi \phi}$ term for simplicity. While it may be comparable in size to $F^+ F^- F^- C^{B \phi} C^{EE} C^{\phi \phi}$ depending on $\ell$, including it may change our value of SNR by at most ${\cal O}(1)$, and thus $\text{SNR} \ll 1$ will still remain.
}
It is then found that the total of these terms simplifies to $F^+ C^{B \phi} \tilde{C}^{BB}$, where $\tilde{C}^{BB}$ denotes the total power spectrum of the unlensed and lensed signals $C^{BB} + F^- F^{-}(C^{E\phi}C^{E\phi} + C^{EE} C^{\phi \phi})$.%
\footnote{We disregard subdominant tensor-mode contributions such as $F^+ F^+ C^{BB} C^{\phi\phi}$ and $F^+ F^+ C^{B\phi} C^{B\phi}$ in $\tilde{C}^{BB}$.
}
A precise computation yields the form of the angle-averaged bispectrum in $\ell_1 + \ell_2 + \ell_3 = {\rm even}$:
\begin{eqnarray}
 B_{{\rm lens}, \ell_1 \ell_2 \ell_3}^{BBB (\rm even)}
  \simeq 
%---------
   F_{\ell_3 \ell_1 \ell_2}^{+}  C_{\ell_1}^{B\phi} \tilde{C}_{\ell_2}^{BB} 
+ 5 \ {\rm perms} ~. \label{eq:BBB_lens}
\end{eqnarray}
This should be regarded as the signal rather than the bias, since this also includes the information on the primary GW. As seen in Fig.~\ref{fig:blllBBB}, this lensed bispectrum can exceed the primary one in high $\ell$. However, its amplitudes are always much smaller than the cosmic variance $\sigma(B_{\ell_1 \ell_2 \ell_3}^{BBB}) \simeq \sqrt{6 \tilde{C}_{\ell_1}^{BB} \tilde{C}_{\ell_2}^{BB} \tilde{C}_{\ell_3}^{BB} }$, and this lensed bispectrum is invisible on our interesting scales. Numerical evaluations in a noiseless all-sky survey lead to $(S/N)_{BBB}^{\rm lens}(\ell_{\rm max} = 300) = 1.4 \times 10^{-3}$ (when $k_* = 7 \times 10^{-5} \, {\rm Mpc^{-1}}$), $6.5 \times 10^{-4}$ (when $k_* = 5 \times 10^{-4} \, {\rm Mpc^{-1}}$), and $8.7 \times 10^{-6}$ (when $k_* = 5 \times 10^{-3} \, {\rm Mpc^{-1}}$). 

While we here skip a curl-mode analysis, which can in principle encode the primary GW signature, it is expected (from Fig.~\ref{fig:ClXp}) that the bispectrum signal related to the curl mode is comparable to the lensing potential one \eqref{eq:BBB_lens} and hence negligible in our bispectrum analysis in Sec.~\ref{sec:bis}.

%%%%%%%%%%%%%%%%%%%%%%%%%%%%%%%%%%%%%%%%%%%%%%%%%%%%%%%%%%%%%%%%%%%%%
\section{Correlations with the lensing potential}\label{appen:pow_lens}

\begin{figure}[t]
  \begin{tabular}{cc}
    \begin{minipage}{0.5\hsize}
  \begin{center}
    \includegraphics[width=1\textwidth]{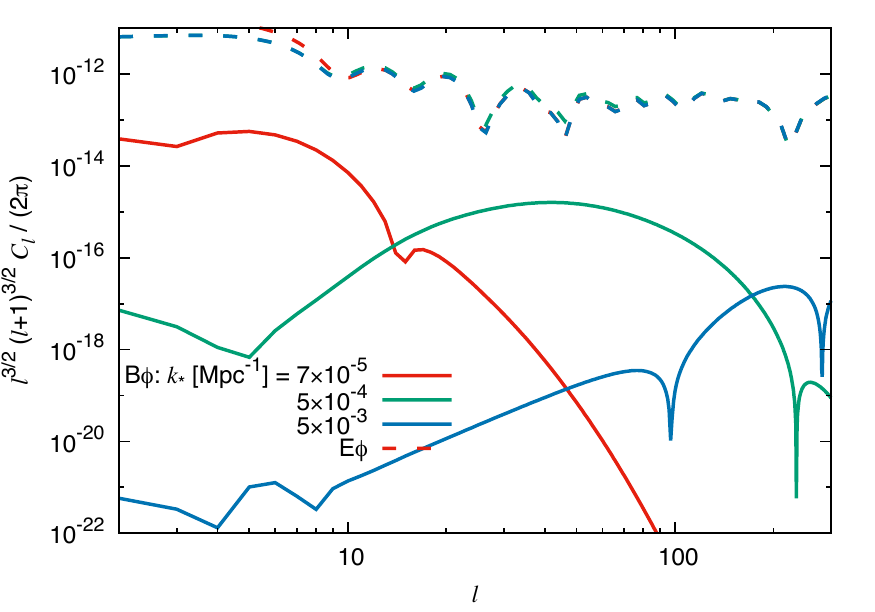}
  \end{center}
\end{minipage}
\begin{minipage}{0.5\hsize}
  \begin{center}
    \includegraphics[width=1\textwidth]{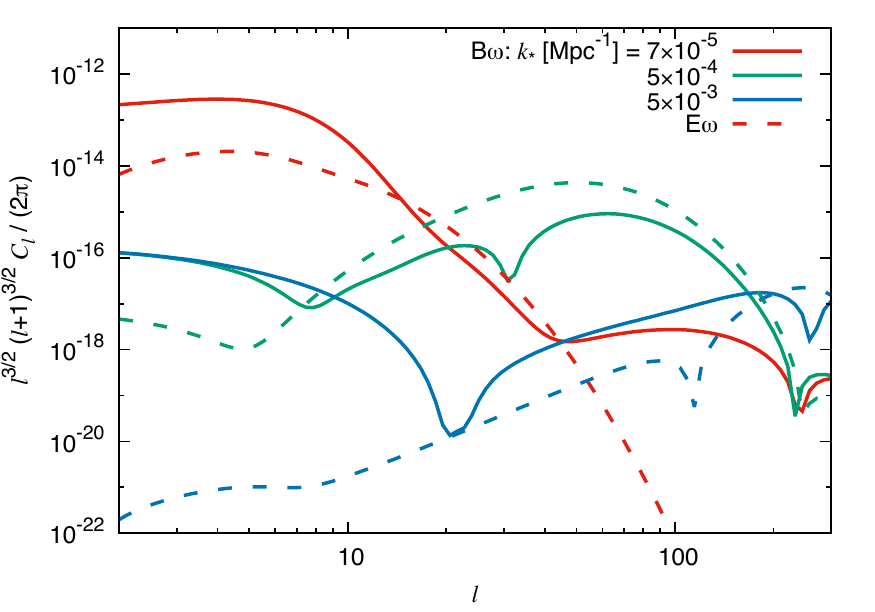}
  \end{center}
\end{minipage}
  \end{tabular}
    \\
    \begin{tabular}{c}
    \begin{minipage}{1.0\hsize}
  \begin{center}
    \includegraphics[width = 0.5\textwidth]{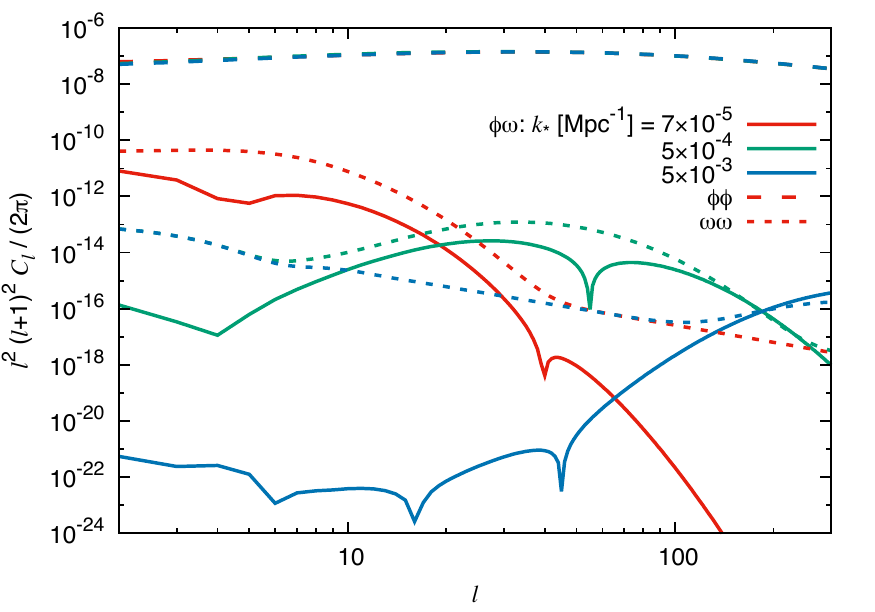}
  \end{center}
\end{minipage}
\end{tabular}
  \caption{Angular correlations of the CMB polarization and the lensing potential, created in the pseudoscalar model.}
\label{fig:ClXp}
\end{figure}

In this Appendix we estimate the power spectra including the lensing potential ($\phi$) and curl mode ($\omega$). 
The harmonic coefficients of the scalar and tensor modes are expressed, respectively, as \cite{Shiraishi:2010sm,Shiraishi:2010kd}
\begin{eqnarray}
    a_{\ell m}^{E (s)} &=& 4\pi i^\ell  \int \frac{d^3 k}{(2\pi)^{3/2}}
 {\cal T}_{E,\ell}^{(s)}(k) \zeta({\bf k}) Y_{\ell m}^*(\hat{k}) ~, \\
%----------
 a_{\ell m}^{\phi (s)} &=& 4\pi i^\ell  \int \frac{d^3 k}{(2\pi)^{3/2}}
 {\cal T}_{\phi,\ell}^{(s)}(k) \zeta({\bf k}) Y_{\ell m}^*(\hat{k}) ~, 
\end{eqnarray}
and 
\begin{eqnarray}
    a_{\ell m}^{E (t)} &=& 4\pi i^\ell  \int \frac{d^3 k}{(2\pi)^{3/2}}
 {\cal T}_{E,\ell}^{(t)}(k) \left[ h_+ ({\bf k}) {}_{-2}Y_{\ell m}^*(\hat{k}) + h_- ({\bf k}) {}_{+2}Y_{\ell m}^*(\hat{k}) \right] ~, \\
%----------
  a_{\ell m}^{B (t)} &=& 4\pi i^\ell  \int \frac{d^3 k}{(2\pi)^{3/2}}
 {\cal T}_{B,\ell}^{(t)}(k) \left[ h_+ ({\bf k}) {}_{-2}Y_{\ell m}^*(\hat{k}) - h_- ({\bf k}) {}_{+2}Y_{\ell m}^*(\hat{k}) \right] ~, \label{eq:almB_ori} \\
%----------
 a_{\ell m}^{\phi (t)} &=& 4\pi i^\ell  \int \frac{d^3 k}{(2\pi)^{3/2}}
 {\cal T}_{\phi,\ell}^{(t)}(k) \left[ h_+ ({\bf k}) {}_{-2}Y_{\ell m}^*(\hat{k}) + h_- ({\bf k}) {}_{+2}Y_{\ell m}^*(\hat{k}) \right] ~, \\
%---------
 a_{\ell m}^{\omega (t)} &=& 4\pi i^\ell  \int \frac{d^3 k}{(2\pi)^{3/2}}
 {\cal T}_{\omega,\ell}^{(t)}(k) \left[ h_+ ({\bf k}) {}_{-2}Y_{\ell m}^*(\hat{k}) - h_- ({\bf k}) {}_{+2}Y_{\ell m}^*(\hat{k}) \right] ~,
\end{eqnarray}
where ${\cal T}_{E/B,\ell}^{(z)}(k)$ and ${\cal T}_{\phi/\omega,\ell}^{(z)}(k)$ are the transfer functions of the CMB polarizations \cite{Zaldarriaga:1996xe,Hu:1997hp} and the lensing potentials \cite{Dodelson:2003bv,Cooray:2005hm,Li:2006si,Yamauchi:2013fra}, respectively, with $z$ denoting the scalar ($z = s$) and tensor ($z = t$) modes. Note that $a_{\ell m}^{B (t)}$ and $a_{\ell m}^{\omega (t)}$ are odd under parity transformation, since they originate from $h_+ - h_-$. These and the expressions of the primordial power spectra \eqref{eq:def-power} result in
\begin{eqnarray}
\Braket{ a_{\ell_1 m_1}^{X_1} a_{\ell_2 m_2}^{X_2} }
  &=& (-1)^{m_1} \delta_{\ell_1, \ell_2} \delta_{m_1, -m_2} C_{\ell_1}^{X_1 X_2} ~,
\end{eqnarray}
with
  \begin{eqnarray}
  C_{\ell}^{\phi\phi}
  &=&
  4\pi  \int_0^\infty \frac{d k}{k}
  \left(
  \left[ {\cal T}_{\phi,\ell}^{(s)}(k) \right]^2 
         %% \nonumber \\ && 
         {\cal P}_\zeta(k)  + 
  \left[ {\cal T}_{\phi,\ell}^{(t)}(k) \right]^2 
         %% \nonumber \\ && 
        {\cal P}_h(k) \right)
         ~, \\
  %----
  C_{\ell}^{\omega\omega}
  &=&  4\pi  \int_0^\infty \frac{d k}{k}
         \left[ {\cal T}_{\omega,\ell}^{(t)}(k) \right]^2 
         %% \nonumber \\ && 
         {\cal P}_h(k) ~, \\
         %----
         C_{\ell}^{\phi\omega}
         &=&  4\pi  \int_0^\infty \frac{d k}{k}
         {\cal T}_{\phi,\ell}^{(t)}(k) {\cal T}_{\omega,\ell}^{(t)}(k)  
         %% \nonumber \\ && 
         {\cal P}_+^{(1)}(k) ~,
  \end{eqnarray}
  and 
  \begin{eqnarray}
    C_{\ell}^{E\phi}
 &=&  4\pi  \int_0^\infty \frac{d k}{k} \left(
 {\cal T}_{E,\ell}^{(s)}(k) {\cal T}_{\phi,\ell}^{(s)}(k)
 {\cal P}_\zeta(k) 
 + {\cal T}_{E,\ell}^{(t)}(k) {\cal T}_{\phi,\ell}^{(t)}(k)
 %% \nonumber \\ && 
 {\cal P}_h(k) \right) ~, \\
 %----
 C_{\ell}^{E\omega}
  &=&  4\pi  \int_0^\infty \frac{d k}{k}
       {\cal T}_{E,\ell}^{(t)}(k) {\cal T}_{\omega,\ell}^{(t)}(k)
         %% \nonumber \\ && 
      {\cal P}_+^{(1)}(k) ~, \\
 %---
 C_{\ell}^{B\phi}
  &=&  4\pi  \int_0^\infty \frac{d k}{k}
       {\cal T}_{B,\ell}^{(t)}(k) {\cal T}_{\phi,\ell}^{(t)}(k)
         %% \nonumber \\ && 
         {\cal P}_+^{(1)}(k) ~, \\
 %----
 C_{\ell}^{B\omega}
  &=&  4\pi  \int_0^\infty \frac{d k}{k}
       {\cal T}_{B,\ell}^{(t)}(k) {\cal T}_{\omega,\ell}^{(t)}(k)
         %% \nonumber \\ && 
         {\cal P}_h(k) ~,
  \end{eqnarray}
where we consider the domain of the model where only the $+$ mode of GWs is sourced, i.e.~$\xi>0$, and therefore neglect the ${\cal P}_-^{(1)}$ contributions, as also done in the main text. We plot these CMB correlations in Fig.~\ref{fig:ClXp}. It is shown that $C_\ell^{E\phi}$ and $C_\ell^{\phi\phi}$ are dominated by the scalar mode and therefore have larger signals than the other correlations sourced by only the tensor mode. The parity-violating correlations $C_\ell^{\phi \omega}$, $C_\ell^{E \omega}$ and $C_\ell^{B\phi}$ arise only from the sourced-mode power spectrum ${\cal P}_+^{(1)}(k)$ and hence they are magnified only at around the peaks due to the sourced mode, i.e., $\ell \sim k_* \tau_0$. $C_\ell^{B\phi}$ is used in computations of the lensed BBB bispectra for $\ell_1 + \ell_2 + \ell_3 = {\rm even}$ \eqref{eq:BBB_lens}.

%%%%%%%%%%%%%%%%%%%%%%%%%%%%%%%%%%%%%%%%%%%%%%%%%%%%%%%%%%%%%%%%%%%%%
\section{Noise spectrum in LiteBIRD}\label{appen:noise}

We here compute the forecast B-mode noise spectrum $N_\ell^{BB}$ in LiteBIRD, used in Secs.~\ref{sec:bis}, \ref{sec:MF}, \ref{sec:comparison} and Appendix~\ref{appen:pow_BB}, by means of \cite{Verde:2005ff,Baumann:2008aq,Oyama:2015gma}. We consider an observation with 15 frequency bands between $40$ and $402 \ \text{GHz}$. The latest information on the LiteBIRD sensitivity \cite{2016JLTP..tmp..169M} is adopted in our computations.

In a practical data analysis, instrumental uncertainties and residual foregrounds due to galactic dust emission and synchrotron radiation reduce sensitivities to primary signals. These effects are quantified by a noise power spectrum \cite{Oyama:2015gma}
\begin{eqnarray}
  N_\ell^{BB} = \left[ \sum_i \frac{1}{n_\ell(\nu_i) + [C_\ell^{S}(\nu_i) + C_\ell^{D}(\nu_i) ]\sigma_{\rm RF} + n_\ell^{\rm RF}(\nu_i)}  \right]^{-1} ~, \label{eq:NlPP_LiteBIRD}
  \end{eqnarray}
where the index $i$ runs over 6 channels for CMB analysis (corresponding to $100 - 235$~GHz \cite{2016JLTP..tmp..169M})  and 
\begin{eqnarray}
 n_\ell(\nu) \equiv \Delta_P^2(\nu) \exp\left[\frac{\ell(\ell+1) \theta_{\rm FWHM}^2(\nu)}{8\ln 2}\right] 
\end{eqnarray}
is determined by instrumental resolutions: the full width at half maximum (FWHM) of a Gaussian beam $\theta_{\rm FWHM}$ (in radian) and dimensionless sensitivities to polarization per $1 \, \rm arcmin^2$ pixel $\Delta_P$, described in \cite{2016JLTP..tmp..169M}. For foreground subtraction, we use 9 channels (corresponding to $40 - 89 \ \text{GHz}$ and $280 - 402 \ \text{GHz}$ \cite{2016JLTP..tmp..169M}). The instrumental uncertainties in this process are expressed as
\begin{eqnarray}
  n_\ell^{\rm RF}(\nu) = \frac{4}{N_{\rm chan}(N_{\rm chan} - 1)} \left[\sum_{j} \frac{1}{n_\ell(\nu_j)}\right]^{-1}
  \left[\left( \frac{\nu}{\nu_{S,\rm ref}} \right)^{2\alpha_S} + \left( \frac{\nu}{\nu_{D,\rm ref}} \right)^{2\alpha_D}  \right] ~,
  \end{eqnarray}
where the index $j$ runs over the 9 ($\equiv N_{\rm chan}$) channels for foreground removal, and $\nu_{S, \rm ref}$ and $\nu_{D,\rm ref}$ are the lowest and highest frequencies in these 9 bands, respectively. The polarization power spectra coming from synchrotron radiation ($C_\ell^{S}$) and dust emission ($C_\ell^{D}$) in our galaxy are modeled, respectively, as \cite{Verde:2005ff,Baumann:2008aq,Oyama:2015gma}
\begin{eqnarray}
  C_\ell^{S} (\nu) &=& {\cal A}_S \left( \frac{\nu}{\nu_{S,0}} \right)^{2\alpha_S}
  \left( \frac{\ell}{\ell_{S,0}} \right)^{\beta_S} ~, \\
  %---
  C_\ell^{D} (\nu) &=& p^2 {\cal A}_D \left( \frac{\nu}{\nu_{D,0}} \right)^{2 \alpha_D}
  \left( \frac{\ell}{\ell_{D,0}} \right)^{\beta_{D}}
\left[ \frac{\exp\left(\frac{h \nu_{D,0}}{k_B T}\right) - 1}{\exp\left(\frac{h \nu}{k_B T}\right) - 1} \right]^2
  ~,
\end{eqnarray}
where each parameter is chosen to be consistent with the results observed in DASI, IRAS, WMAP and {\it Planck}: ${\cal A}_S = 6.3 \times 10^{-18}$, $\alpha_S = -3$, $\beta_S = -2.6$, $\nu_{S,0} = 30 \, {\rm GHz}$, $\ell_{S,0} = 350$, ${\cal A}_D = 1.3 \times 10^{-13}$, $\alpha_D = 2.2$, $\beta_{D} = -2.5$, $\nu_{D,0} = 94 \, {\rm GHz}$, $\ell_{D,0} = 10$, $T = 18 \, \rm K$ and $p = 0.15$. In Eq.~\eqref{eq:NlPP_LiteBIRD}, $\sigma_{\rm RF}$ expresses the percentage of residual foreground in CMB maps, and we assume $\sigma_{\rm RF} = 4 \times 10^{-4}$, 
corresponding to 2\% level \cite{Katayama:2011eh}. 

We have included the noise spectrum \eqref{eq:NlPP_LiteBIRD} in Fig.~\ref{fig:ClBB_pseudo}, showing that, for $\ell \lesssim 20$ and $\ell \gtrsim 200$, it exceeds the lensed BB spectrum due to high contamination by residual foreground and the lack of instrumental resolution, respectively. 

%########################################
% Create the reference section using BibTeX
\bibliography{paper}
%\nocite{*}
%%%%%%%%%%%%%%%%%%%%%%%%%%%%%%%%%%%%%%%%%%

\end{document}